\documentclass[journal, 10pt]{IEEEtran}
\usepackage[utf8]{inputenc}
\usepackage{graphicx}
\usepackage{xcolor}
\usepackage{multirow}

\usepackage{float}
\floatstyle{plaintop}
\restylefloat{table}
\usepackage{hyperref}
\usepackage{multirow}

\title{deSpeckNet: Generalizing Deep Learning Based SAR Image Despeckling.  \thanks{\noindent Adugna G.~Mullissa, Diego Marcos,  Martin Herold and Johannes Reiche are with the Laboratory of Geo-information Science and Remote sensing, Wageningen University, 6700 AA  Wageningen, The Netherlands.  (email: adugna.mullissa@wur.nl, diego.marcos.gonzalez@wur.nl, martin.herold@wur.nl and johannes.reiche@wur.nl). Devis Tuia is with the ECEO lab in the Swiss Federal Institute of Technology (EPFL)-Sion (email:devis.tuia@epfl.ch) }}
\author{Adugna G. Mullissa, Diego Marcos, Devis Tuia, Martin Herold and Johannes Reiche }

\newcommand{\diego}[2]{\textcolor{blue}{#2}}

\begin{document}

\maketitle

\begin{abstract}
\noindent  Deep learning (DL) has {proven to be a suitable approach for} despeckling synthetic aperture radar (SAR) images. So far, most  DL models are {trained to reduce speckle that follows a particular distribution, either using simulated noise or a specific set of real SAR images}, limiting the applicability of these methods for real SAR images {with unknown noise statistics}. In this paper, we present a DL method, deSpeckNet\footnote{\url{https://github.com/adugnag/deSpeckNet}}, that estimates the speckle noise {distribution} and the despeckled image simultaneously. Since it does not depend on a specific noise model, deSpeckNet generalizes well across SAR acquisitions in a variety of landcover conditions.  We evaluated the performance of deSpeckNet on single polarized Sentinel-1 images acquired in Indonesia, The Democratic Republic of Congo and The Netherlands, a single polarized ALOS-2/PALSAR-2 image acquired in Japan and an Iceye X2 image acquired in Germany. In all cases, deSpeckNet was able to effectively reduce speckle and restore the images in high quality with respect to the state of the art.\\
\end{abstract}


\IEEEpeerreviewmaketitle

\section{Introduction}

\noindent The recent availability of global Earth observation SAR data, for instance from the Sentinel-1 SAR satellites, has been a game changer for large scale, all weather, day/night monitoring of land surfaces. However, the applicability of these datasets has been limited by the presence of speckle. Speckle is inherent in all SAR images as they are acquired by a coherent active microwave imaging system. Speckle occurs when the backscattered signal from independent targets is coherently superimposed within each resolution cell. Depending on the size of the resolution cell, the superimposition of these signals results in interference whose effect is observed as speckle. Speckle is a true scattering information collected from a target, but for image processing purposes it is often considered as noise. To improve the radiometric quality of SAR images before analysis, speckle has to be reduced.  In fact, speckle reduction has been an active research area since the advent of airborne and space-borne imaging SAR sensors in the 1970s \cite{porcello1976speckle}. \\

\begin{figure}
	\centering
	\includegraphics[width=0.5\textwidth]{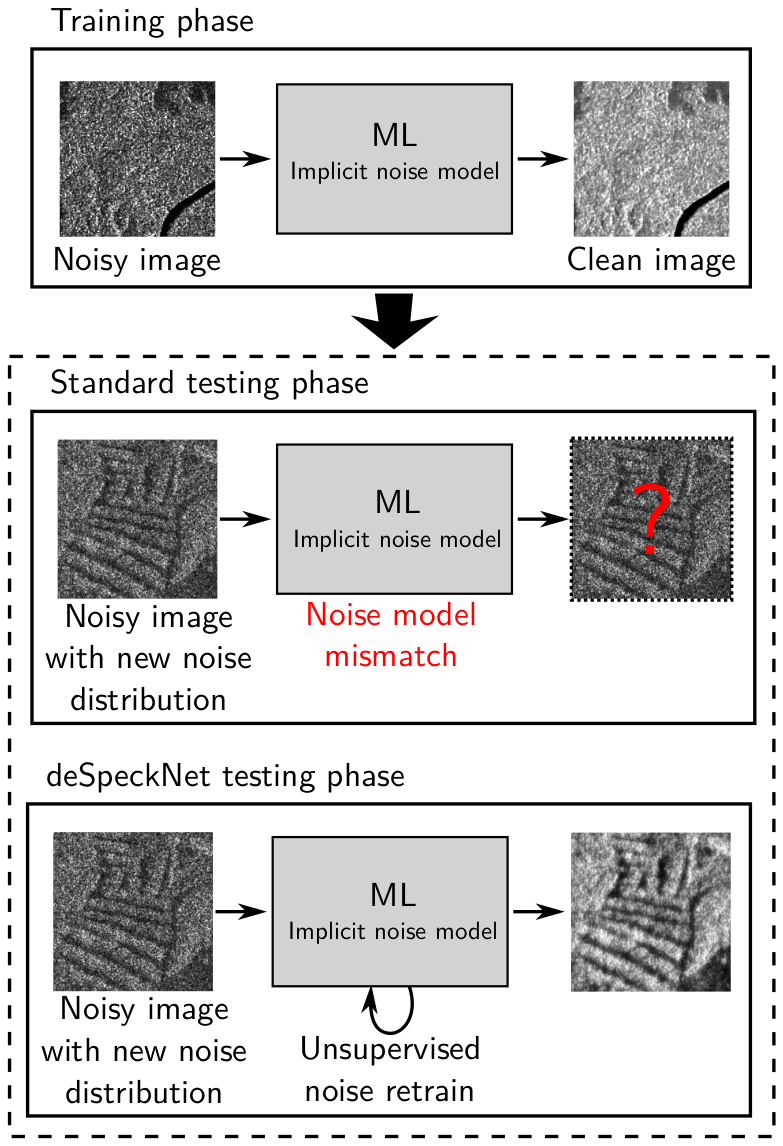}
	\caption{When a Machine Learning method is applied to image denoising, it implicitly learns a noise model from the training examples. If there is a mismatch between this distribution and that of the test images the performance can drop substantially. In this work we propose a method to readjust the noise model to the test images without any additional clean ground truth images.}
	\label{fig:Fig1}
\end{figure}

\noindent Earlier speckle filtering methods focused on spatially adaptive filters. Most of these were  based on pixel intensity statistics determined on a local neighbourhood window.  These class of filters operated in a sliding window fashion, where the pixel to be filtered is the center pixel in the moving window. The boxcar filter \cite{lee1980digital} is the simplest spatial filter that estimates the mean value of all the pixels in the moving window. The boxcar filter was effective in reducing speckle in homogeneous regions at the cost of resolution.  The Lee filter \cite{lee1994intensity} {reduced the impact of} the loss in resolution by estimating the minimum mean square error in a neighbouring window. The Frost filter \cite{frost1982model} applied exponential weighting and damping factor to control the amount of filtering in a low pass filtering setting. These methods improved the preservation of features in speckle filtering, however, they introduced artefacts along feature boundaries.  
To overcome this problem, the authors in \cite{lee1999polarimetric} proposed to select similar pixels by using a series of edge aligned non-rectangular windows, \cite{vasile2006intensity} used intensity-driven neighborhood region growing based on the image intensity and \cite{lee2005scattering} proposed selecting similar neighboring pixels based on scattering characteristics. {In addition}, \cite{deledalle2014nl} used non-local means with weighted maximum likelihood estimation to reduce speckle, and the 3-D block matching approach (BM3D) \cite{dabov2007image}, {which} groups image patches into 3-D arrays based on their similarity and performs estimations into a 2-D image array from the grouped blocks.\\

\noindent A second family of approaches exploits the wavelet transform of the single look image in log form. Notable works involve the wavelet Bayesian denoising that is introduced by \cite{molina2011evaluation}, based on Markov Random Fields (MRF). Authors in \cite{mahdianpari2017effect} introduced a  SAR image despeckling method that is based on adaptive Gauss-MRF. In \cite{lopes1990maximum}, a homomorphic wavelet maximum a-posteriori (MAP) filter was introduced improving the performance of the original Gamma-MAP speckle filter. \\

\noindent A third family of approaches has recently started to attract attention: thanks to the advent of powerful computation capability, significant advances have been made using deep learning (DL) methods to perform image de-noising tasks. The most notable difference between these methods and those described earlier is that DL based methods { \emph{learn} a suitable de-noising function based on pairs of noisy and clean images, instead of using a pre-defined function}. \cite{chierchia2017sar} implemented SAR-CNN by adopting the concept of residual learning and deep Convolutional Neural Network (CNN) proposed by \cite{zhang2017beyond} for additive white Gaussian noise reduction. In SAR-CNN, the input SAR images are transformed to the homomorphic form and used for training the SAR-CNN network. The network uses a temporally averaged image as clean reference label, \emph{i.e.} as a proxy of the speckle free reference image. Once the network is trained, the prediction image is transformed back to the original image domain by using an exponential function. The authors in \cite{zhang2018learning} used a dilated residual network (DRN) using skip connections to train a deep neural network for SAR image de-speckling. 
The marked difference between SAR-CNN and SAR-DRN was the usage of real SAR images in SAR-CNN, whereas SAR-DRN 1) was trained on simulated images 2) exploited  residual connections 3) processed images in their native form. Recent works focused on combining  loss functions with different purposes: for example, \cite{vitale2019new} uses simultaneously a mean square error (MSE) loss that reconstructs the noise free image and a Kulback-Lieber loss to reconstruct the distribution of the speckle. Furthermore, authors in \cite{pan2019filter} deal with the unknown noise statistics in SAR images by embedding a CNN model for additive white Gaussian noise reduction with a Multi-channel Logarithm with Gaussian denoising (MuLoG) algorithm  for multiplicative noise, first {introduced} in \cite{deledalle2017mulog}.  \\

\begin{figure*}
    \centering
    \begin{tabular}{cc}
    \includegraphics[width=0.5\textwidth]{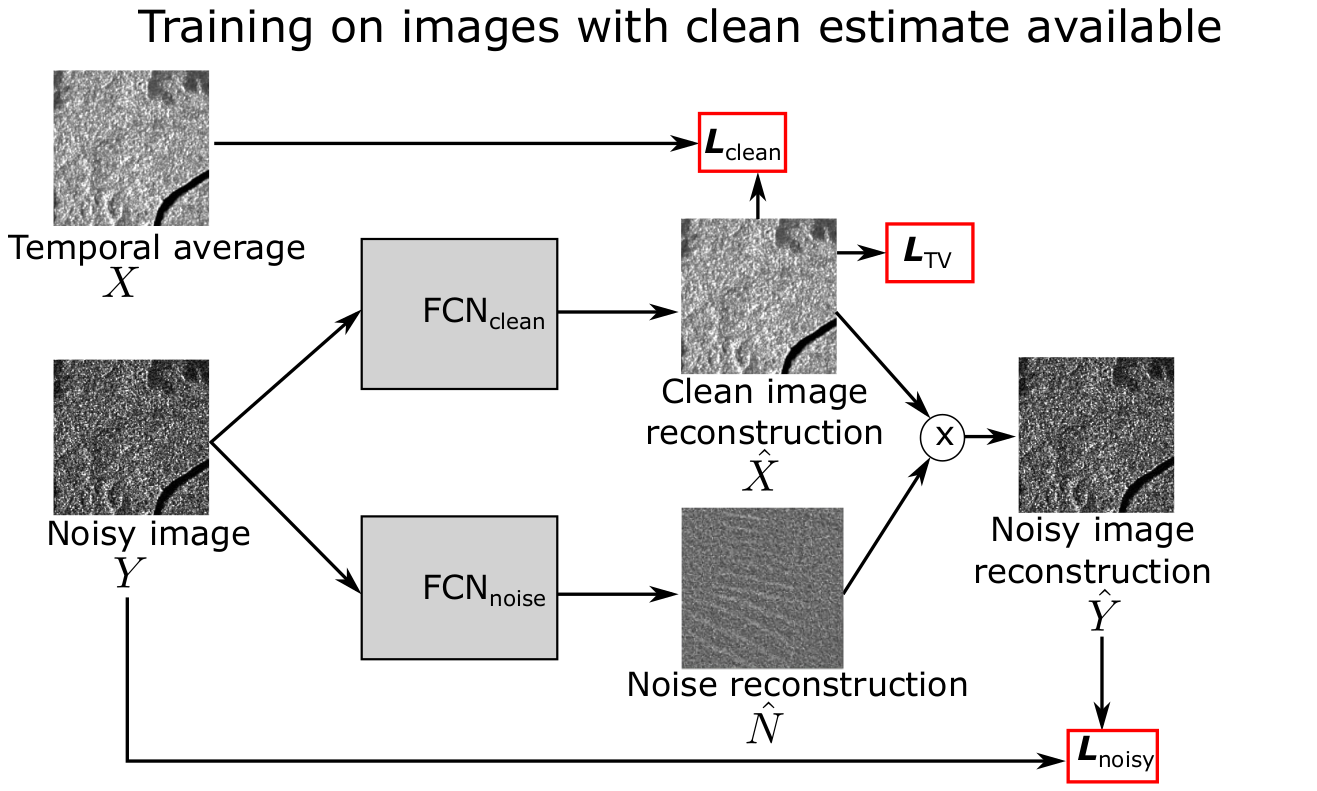} &
    \includegraphics[width=0.5\textwidth]{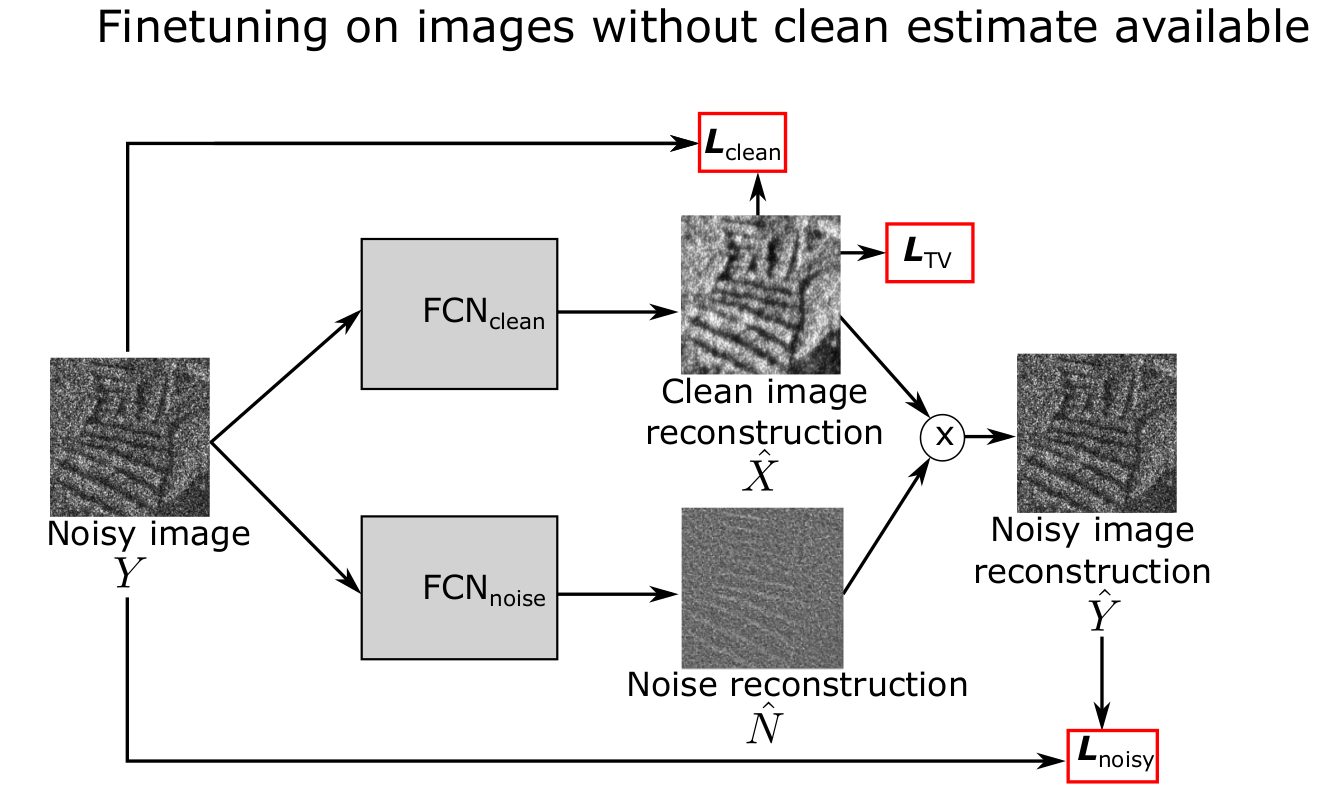}\\
    (a) Phase 1 & (b) Phase 2
    \end{tabular}
    \caption{The two phases of deSpeckNet. (a) The model is first trained with an image for which a clean estimate exists (\emph{e.g.} a temporal average). Three losses are used: an MSE based loss $L_{Clean}$, $L_{TV}$ and $L_{Noisy}$  reconstruction losses, for the clean and the noisy (original) images, respectively (b) When moving to an image for which no clean reference is available, an MSE based loss $L_{Clean}$ and a total variation loss $L_{TV}$ on the clean image reconstruction and an MSE based loss $L_{Noisy}$ for the noisy image reconstruction.}
    \label{fig:Fig2}
\end{figure*}

\noindent In a supervised learning paradigm, the quality of the prediction depends on the quality of the reference labels used for training. {For} generating the reference labels, two approaches have been followed in the literature. The first is the model based simulation from known speckle statistics \cite{lattari2019deep}. This approach relies on artificially simulating the speckle noise and adding it to an optical image (referred to as a clean image) by following an additive white Gaussian noise model \cite{zhang2017beyond} or a multiplicative noise model \cite{lattari2019deep}. This approach has three major drawbacks:\\
\begin{enumerate}
\item {It requires to} assume an a-priori speckle noise model based on a Gamma probability distribution and a multiplicative noise model. This poses a problem for adapting the network to data that does not follow the same noise model. A good example for this is the difference in the noise statistics between SAR images with different resolution, band and landcover types, such as the intensity images of high resolution single look Iceye X2 X-Band SAR images and medium resolution pre-processed and multi-looked Sentinel-1 C-Band Ground range detected (GRD) images. Therefore, a model trained on the single look SAR data {would not} necessarily adapt to a pre-processed SAR image, as illustrated in Figure~\ref{fig:Fig1}. Hence, it is imperative to design a model that is robust to changes in speckle noise statistics. \\

    \item The artificial simulation and addition of noise to an optical image does not represent the true appearance of real SAR images. This is exemplified in the representation of deterministic scatterers in the SAR images.  {This phenomenon is not captured by an artificial simulation of noise in an optical image.} Hence, a model {trained} on these simulated \diego{trained}{} images {can not} recognize these features when applied to a true SAR image.\\
    
    \item  The white noise property of the simulated images does not represent the scatterer dependent spatial distribution of the noise in semi-heterogeneous and heterogeneous media. Hence, a network trained on homogeneous noise statistics tends to over smooth features in a heterogeneous scenes, resulting in sub-optimal results.  \\
\end{enumerate}

\noindent {These problems led to the second approach: to use real SAR images as labels. However, to obtain a noise-free training label for real SAR images is impossible, because speckle is inherent in all SAR images. One solution proposed in the literature is selecting an area where there is a multi-temporal image stack with as little temporal change as possible and taking a temporal average to estimate a noise-free image to be used as a {proxy for the} reference label. This approach has been demonstrated to provide good results \cite{chierchia2017sar}. However,  when the scene under investigation is  non-stationary in time, taking the temporal average will result in an erroneous estimation of the reference label, limiting the applicability of this method. Hence, it is imperative to design a network that can generalize well in areas where a temporal average label is not available.}\\

\noindent In this paper, we propose a deep learning {pipeline} named deSpeckNet {that can despeckle SAR images with unknown noise statistics}. Initially, deSpeckNet is trained using a temporally averaged SAR image as a reference label. In this first step, the model simultaneously estimates the noise free image and the noise {component}.  In the second step, {this model is subsequently} fine-tuned to fit any type of SAR image acquired over any type of cover conditions without using any clean reference labels, \emph{i.e.} in an unsupervised way. To show the versatility of the approach, the proposed method is evaluated on SAR images acquired in Indonesia, the Democratic Republic of Congo (DRC), The Netherlands, Japan and Germany and across several SAR sensor configuration involving different noise models. \\

\noindent The paper is organised as follows. Section II describes the proposed methodology. The datasets used  are described in Section III. Section IV describes the experimental settings, whereas  Section V provides results and  discussions. Conclusions are presented in Section VI.\\

\section{deSpeckNet}

\noindent In a distributed {medium}, a SAR image {with a fully developed speckle} is assumed to follow a multiplicative speckle model \cite{ulaby2014microwave}:

\begin{equation}
    Y = X N,
\end{equation}

\noindent where $Y$ is the observed SAR intensity image, $X$ is the {underlying radar reflectivity of the scene, which can be viewed as hypothetically noise free intensity image (since SAR images can not exist without speckle)}  and $N$ is the speckle {image}. The random speckle noise follows a Gamma probability density function:

\begin{equation} \label{eq:2}
    p(N|L) = \frac{1}{\Gamma(L)} L^L N^{L -1} e ^{{-N}{L}}, ~~~N \geq 0, L \geq 1,
\end{equation}

\noindent where $\Gamma$ is a Gamma function and $L$ is interpreted as the number of looks of the SAR image. For single look SAR image ($L=1$),  (2) simplifies to an exponential distribution. If the SAR image is in the amplitude domain, N is characterized by a Nakagami probability density function: \\

\begin{equation}
    p(N|L) = \frac{1}{\Gamma(L)} 2 L^L N^{2L-1} e ^{{N^2}{-L}}, ~~~N \geq 0, L \geq 1,
\end{equation}

\noindent In the single look amplitude case, (3) simplifies to the Rayleigh probability density function. \\

\noindent The proposed deep learning based de-speckling method (deSpeckNet)  {consists of two phases:} the first follows a supervised learning paradigm by using a temporally averaged SAR image as a reference label. The second phase {consists of} unsupervised fine-tuning \diego{}{} that learns to adapt to a new noise distribution. To this end, we design the architecture of deSpeckNet to follow the multiplicative noise model defined in (1).\\

\subsection{Architecture}

\noindent We use a Siamese architecture to estimate the noise free image ($\hat{X}$) and the estimated noise ($\hat{N}$)  separately (Figure 2). We adopted this architecture to provide the possibility to tune the network to the noise statistics of other SAR images from a different region or from a different sensor. 
The two identical branches estimate the clean image $\hat{X}$ ($\mathrm{FCN_{clean}}$) and the noise $\hat{N}$ ($\mathrm{FCN_{noise}}$). With those two components, we reconstruct the input noisy image ($\hat{Y}$) using (1). \\

\noindent Both ($\mathrm{FCN_{noise}}$) and ($\mathrm{FCN_{clean}}$) consist of four main building blocks namely convolution \cite{mullissa2019polsarnet}, batch normalization \cite{ioffe2015batch}, non-linear activation \cite{dahl2013improving} and loss function. The architecture for deSpeckNet {does not} use any pooling layers to avoid using \diego{the required}{} upsampling layers to reconstruct the images to their original sizes \cite{lattari2019deep}, as these lay additional computational burden. Instead, we opted to maintain the sizes of feature maps in the intermediate layers and increase the depth of the network. \\


\noindent  To train the network, we apply three types of loss functions, the mean square error-based $L_{Clean}$, $L_{Noisy}$ losses and a total variation loss $L_{TV}$, combined as:

\begin{equation}
    Loss = L_{Clean} + L_{Noisy} +L_{TV}
\end{equation}

\noindent In the first, supervised training phase (Figure~\ref{fig:Fig2}a), we apply a mean square error (MSE) based $L_{Clean}$ loss between the clean label {$X$} and reconstructed clean image {$\hat{X}$}:
\begin{equation}
L_{Clean}(\mathbf{X},\hat{\mathbf{X}}) = \mu \frac{1}{n} \sum_{i=1}^n (X_i - \hat{X}_i)^2,
\end{equation}
\noindent where $n$ is the number of pixels in a training patch and $\mu$ is the weight assigned to the loss.
Once $\hat{N}$ and $\hat{X}$ \diego{counterparts}{} are reconstructed in the network, we apply an element-wise multiplication following the multiplicative noise model used for SAR images (1). The reconstructed noisy image $\hat{Y}$ is finally compared to the input noisy SAR image {$Y$} using another MSE loss, $L_{Noisy}$ (Figure~\ref{fig:Fig2}):
\begin{equation}
L_{Noisy}(\mathbf{Y},\hat{\mathbf{Y}}) = \xi \frac{1}{n} \sum_{i=1}^n (Y_i - \hat{Y}_i)^2.
\end{equation}
Here, $\xi$ is the weight assigned to the loss. The usage of {this} second loss function is important for providing a learning signal {in the second phase, when} no temporal average is available. This approach makes deSpeckNet different from the other deep learning based approaches for denoising SAR images. \\

\noindent {In the second unsupervised fine-tuning phase (Figure~\ref{fig:Fig2}b)}, since a temporally averaged image is assumed to be unavailable, we use the input noisy image as a reference label and {down-}weight  the $L_{Clean}$ loss by a small value $\mu_2$:
\begin{equation}
L_{Clean,2}(\mathbf{Y},\hat{\mathbf{X}}) = \mu_2 \frac{1}{n} \sum_{i=1}^n (Y_i - \hat{X}_i)^2.
\end{equation}
This has the effect of maintaining the solution close to the original image so that spatial structures are preserved while the other losses smooth ($L_{TV}$) and de-noise ($L_{Noisy}$). \\

\noindent {For smoothing, w}e used {a Total Variation~\cite{aubert2008variational} (TV) loss}  $L_{TV}$ {in order to encode a smoothness prior on the clean image:}
\begin{equation}
    L_{TV}(\mathbf{\hat{X}}) = \lambda \sum_i |\nabla \hat{X_i}|.
\end{equation}
{The $L_{TV}$ loss minimizes the absolute differences between neighboring pixel-values, enforcing smoothness while preserving edges.}
\\

\section{Datasets}

\noindent To evaluate the performance of deSpeckNet we used a Sentinel-1 ground range detected (GRD) image time series with 23 images acquired over Pegunungan Barisan, Sumatra, Indonesia, to synthesize the reference labels and train the initial model. {Since  deSpeckNet is designed to adjust to different noise levels in different SAR images, the decision to  train the model based on Sentinel-1 GRD is based  on convenience for synthesising a reference label image with fewer number of images. } To demonstrate the performance of deSpeckNet in despeckling SAR images without a temporally averaged label, {and obtained using different sensors and across multiple regions, we fine-tune the model on}:\\

\begin{enumerate}
    \item {Images from} different geographic region and landcover type, we used three study areas, each are composed of a single Sentinel-1 GRD image acquired over the Kindu area in the Democratic Republic of Congo (DRC), the city of Utrecht and the region of Flevoland in the Netherlands, respectively. \\
    
    \item {Images} from different sensors and landcover types. We used an ALOS-2/PALSAR-2 image acquired over Fujiyama, Japan and an Iceye X2 image acquired near the city of Kiel in Germany.
\end{enumerate}

 \diego{These images are all selected over natural environments as the contrast between ground features is very low and the speckle level is almost on the same level as the desired signal. To evaluate the performance of deSpeckNet over urban regions, we used a sentinel-1 image acquired over the city of Utrecht in the Netherlands.}{}
 
 \begin{table*}[tbh]
\label{tab:prop_FCN}
\begin{center}
\footnotesize
\begin{tabular}{lllll}
\hline Parameter & Indonesia & DRC & Utrecht & Flevoland\\ 
\hline Polarization & VH & VH & VV &  VH\\
Product& GRD & GRD & GRD & GRD\\
Acquisition mode & IW & IW & IW & IW \\
Resolution  &  $10\mathrm{m}\times 10\mathrm{m}$ & $10\mathrm{m}\times 10\mathrm{m}$ &    $10\mathrm{m}\times 10\mathrm{m}$ &
$10\mathrm{m}\times 10\mathrm{m}$\\ 
Incidence angle & $40.2^0$ & $41.2^0$ & $40.8^0$ & $40.8^0$\\   
Orbit &   Descending & Descending & Descending & Descending \\ 
Dates &  July 05, 2018 & August 26, 2017 &October 11, 2018 & October11, 2018  \\ 
& April 19, 2019 & August 9, 2018 & July 2, 2019 & -\\
Multi-temporal images &  23 & 29 & 22 & 1 \\  \hline
\end{tabular}
\end{center}
\caption{Acquisition parameters for the Sentinel-1 training and test images.}\label{tbl:Tab11}
\end{table*}

\subsection{Sentinel-1 data}

\noindent  {The Sentinel-1 GRD images used in the experiments are acquired in the interferometric wide swath (IW) mode with a technique known as TOPS (Terrain Observation  with Progressive Scan). They were acquired in C-Band for both single and dual polarization. The GRD images were multi-looked to 5 looks in the range direction and projected to ground range using an Earth ellipsoid model by the data provider.} The Sentinel-1 datasets used in this paper are acquired from four regions (Table~\ref{tbl:Tab11}):

\begin{enumerate}
    \item \textit{Indonesia}: The training image for deSpeckNet was acquired over the Pegunungan Barisan area in Sumatra, Indonesia. It consisted of an image with $1682 \times 2300$ pixels. The  multi-temporal images used to synthesize the training labels were acquired from July 5, 2018 to April 19, 2019. To assess the performance of deSpeckNet in tuning the network for a different region, we used a mono-temporal images acquired on July 5, 2017. The area is mostly covered by oil palm plantations and forests. There are hardly any urban region within the area.
    
    \item \textit{DRC}:  To assess the performance of deSpeckNet in tuning the network for a different region, we used a Sentinel-1 image acquired over Kindu in the DRC. The image was acquired on August 26, 2017 and it consists of an image with $1001 \times 1001$ pixels. This test area is mostly covered by primary forest with some bare soil. To synthesize the clean reference label for assessing the performance of deSpeckNet, we used 29 multi-temporal images acquired over the same area from August 26, 2017 to August 9, 2018. 
    
    \item \textit{Netherlands-Utrecht}: To assess the performance of deSpeckNet in tuning the network for a different region and landcover type, we used a Sentinel-1 image acquired over the city of Utrecht in the Netherlands. The image scene is dominated by urban areas. The images were acquired on October 11, 2018 consisting of $1,360 \times 2,087$ pixels. These images were selected to demonstrate the generalization capability of deSpeckNet in urban regions. We also used 22 multi-temporal images acquired from October 11, 2018 to July 2, 2019 to evaluate the performance of deSpeckNet quantitatively.
    
    \item   \textit{Netherlands-Flevoland}: To demonstrate the performance of deSpeckNet in a temporally unstable region, we used a Sentinel-1 image acquired over the Dutch region of Flevoland. The image was acquired on October 11, 2018 and is $1,821 \times 1,204$ pixels wide. Since it is an agricultural area, the temporally stable backscatter assumption {could not} be fulfilled to synthesize the reference labels and this case is assessed only qualitatively. 
\end{enumerate}

\begin{table}[tbh]
\label{tab:prop_FCN}
\begin{center}
\footnotesize
\begin{tabular}{lll}
\hline Parameter & ALOS-2 & Iceye\\ 
\hline Polarization  & VH & VV \\
Sensor&  PALSAR-2 &  X2\\
Product&  L1.5 & SLC\\
Acquisition mode & Strip map & Strip map\\
Resolution & $2.5\mathrm{m}\times 2.5\mathrm{m}$ &    $0.71\mathrm{m}\times 1.48\mathrm{m}$   \\ 
Incidence angle & $23.6^0$ & $20.4^0$\\   
Orbit  & Descending & Descending \\ 
Dates  & June 06, 2014 &April 29, 2019  \\ 
Multi-temporal images &  1 & 1 \\  \hline
\end{tabular}
\end{center}
\caption{Acquisition parameters for the ALOS-2-PALSAR-2 and Iceye X-2 test images.}\label{tbl:Tab10}
\end{table}

\subsection{ALOS-2/PALSAR-2 image}

\noindent {The ALOS2-PALSAR2 image is acquired in stripmap mode and is also multilooked 2 times in the azimuth direction and is projected to ground range using an Earth ellipsoid model by the data provider.} The ALOS-2/PALSAR-2 sensor acquires data in L band for both single, dual and and quad polarization data (Table~\ref{tbl:Tab10}).  The mono-temporal test image is acquired over the Fujiyama area in Japan on June 06, 2014. It consists of an images image with $1,060 \times 1,601$ pixels. The area is a natural environment covered by forests and some bare areas. We selected this sensor and image to demonstrate the performance of deSpeckNet in adapting to a new geographic region and new sensor. It was not possible to freely acquire a multi-temporal images to synthesize the temporally averaged labels for quality evaluation, as it is a commercial sensor. Therefore, and as in the Flevoland case, we assess the results only qualitatively.\\

\subsection{Iceye X2 image}

\noindent The Iceye X2 sensor acquires data in X band for single polarization data {in Strip map mode} (Table~\ref{tbl:Tab10}).  The image is acquired near the city of Kiel in Northern Germany and is $1,287 \times 958$ pixels. The area is a mixed scene of agricultural and urban region. We selected this region to demonstrate the performance of deSpeckNet when applied to a high resolution image in a different region and landcover type. In this scene, it was also not possible to freely acquire a multi-temporal images to synthesize the temporally averaged labels for quality evaluation, as it is a commercial sensor.\\

\begin{figure*}
	\centering
	\begin{tabular}{ccc}
		\includegraphics[width=5cm,height=4cm]{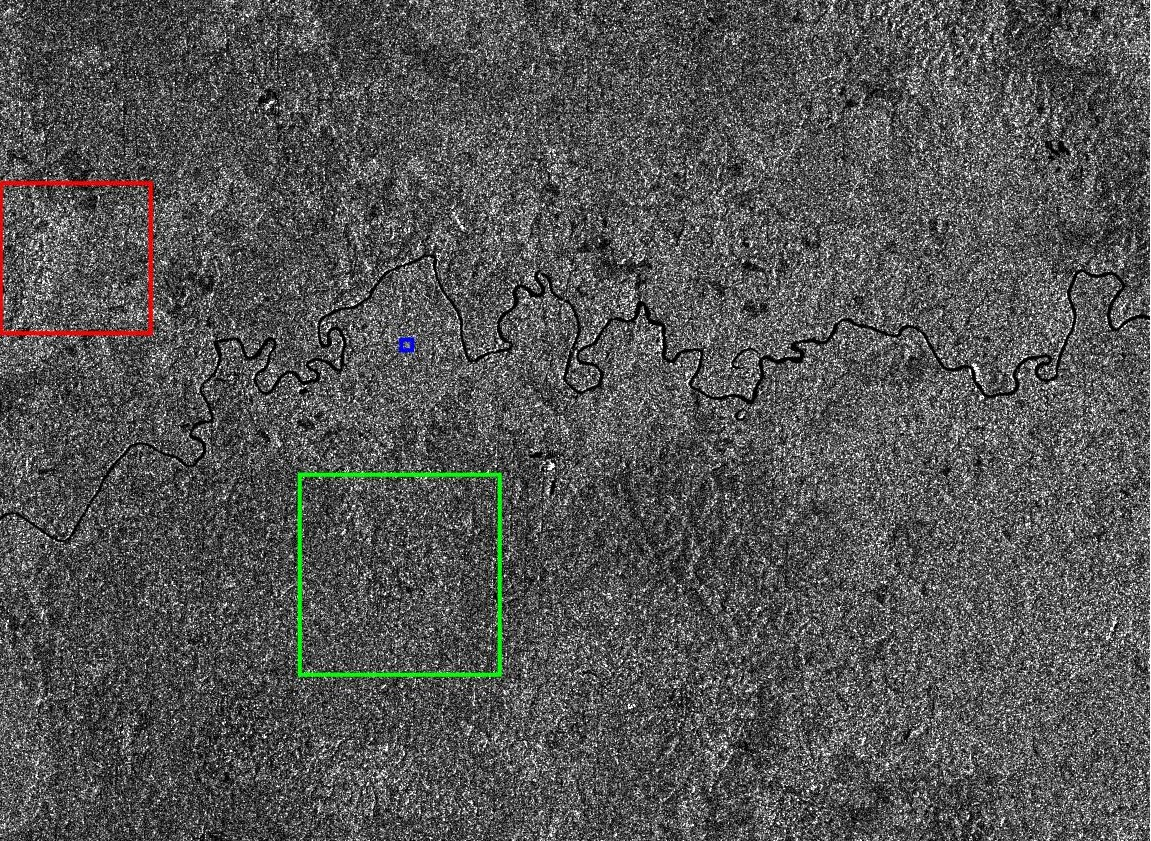}&
		\includegraphics[width=5cm,height=4cm]{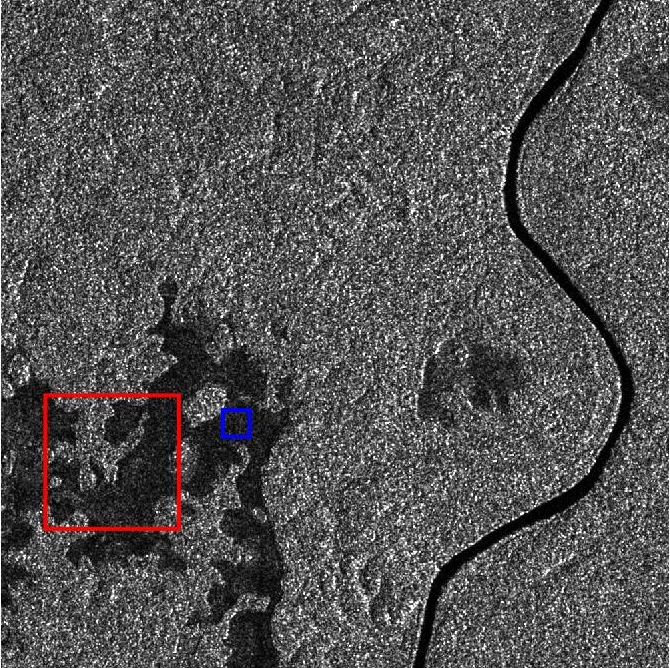} &
		\includegraphics[width=5cm,height=4cm]{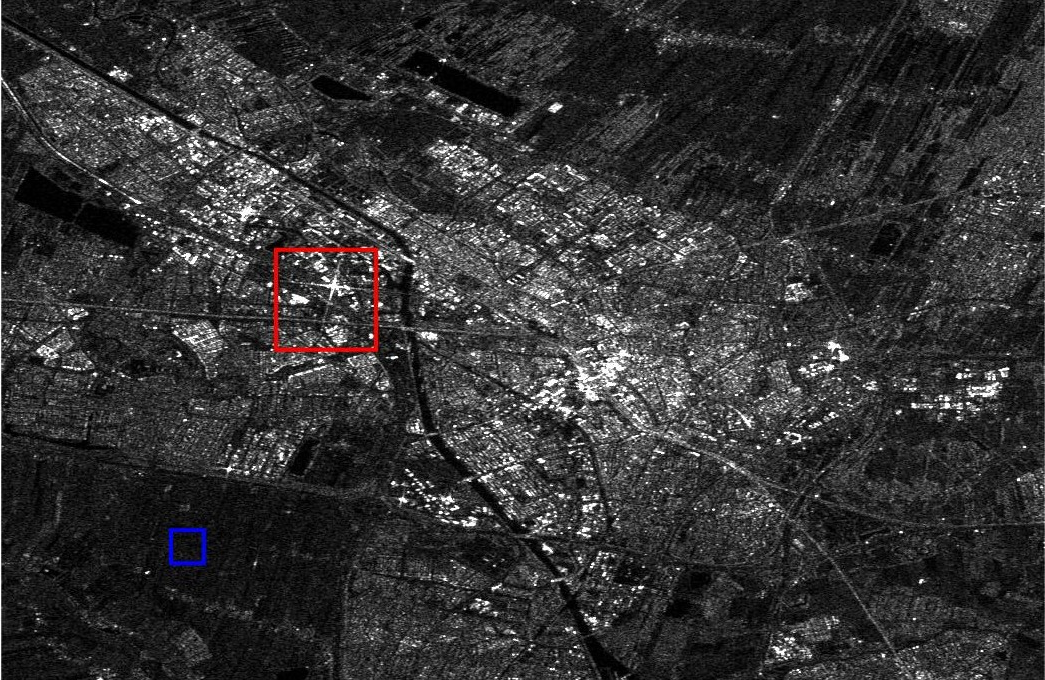} \\
		{(a)}&{(b)} &{(c)}
	\end{tabular}
	\begin{tabular}{ccc}
		\includegraphics[width=5cm,height=4cm]{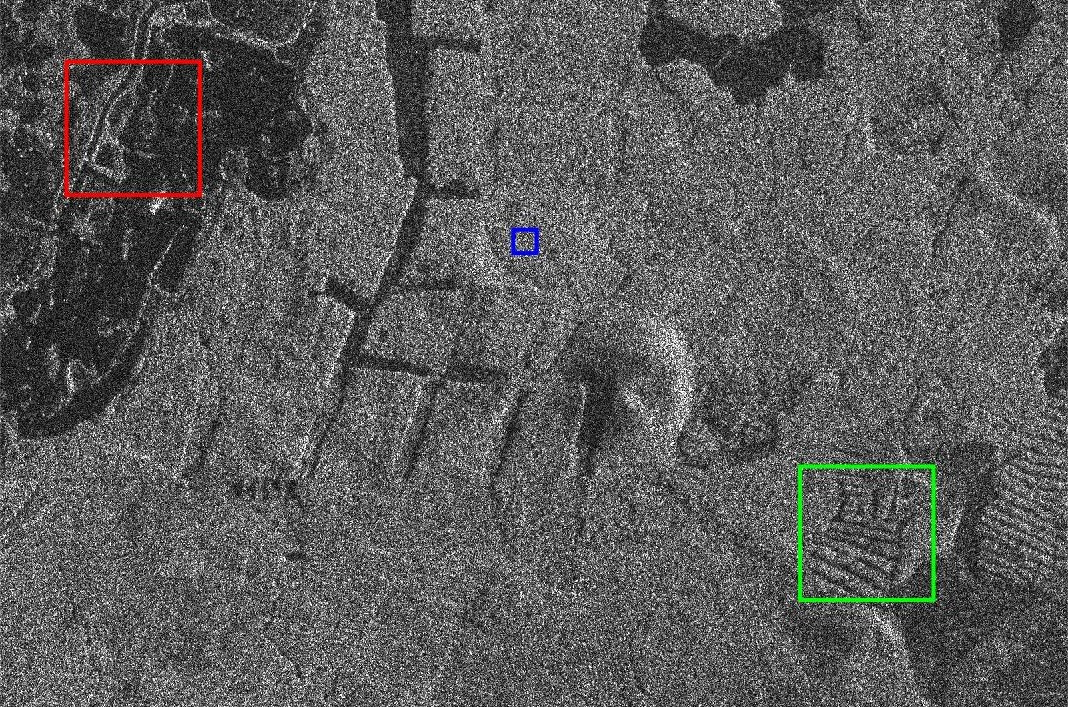}&
		\includegraphics[width=5cm,height=4cm]{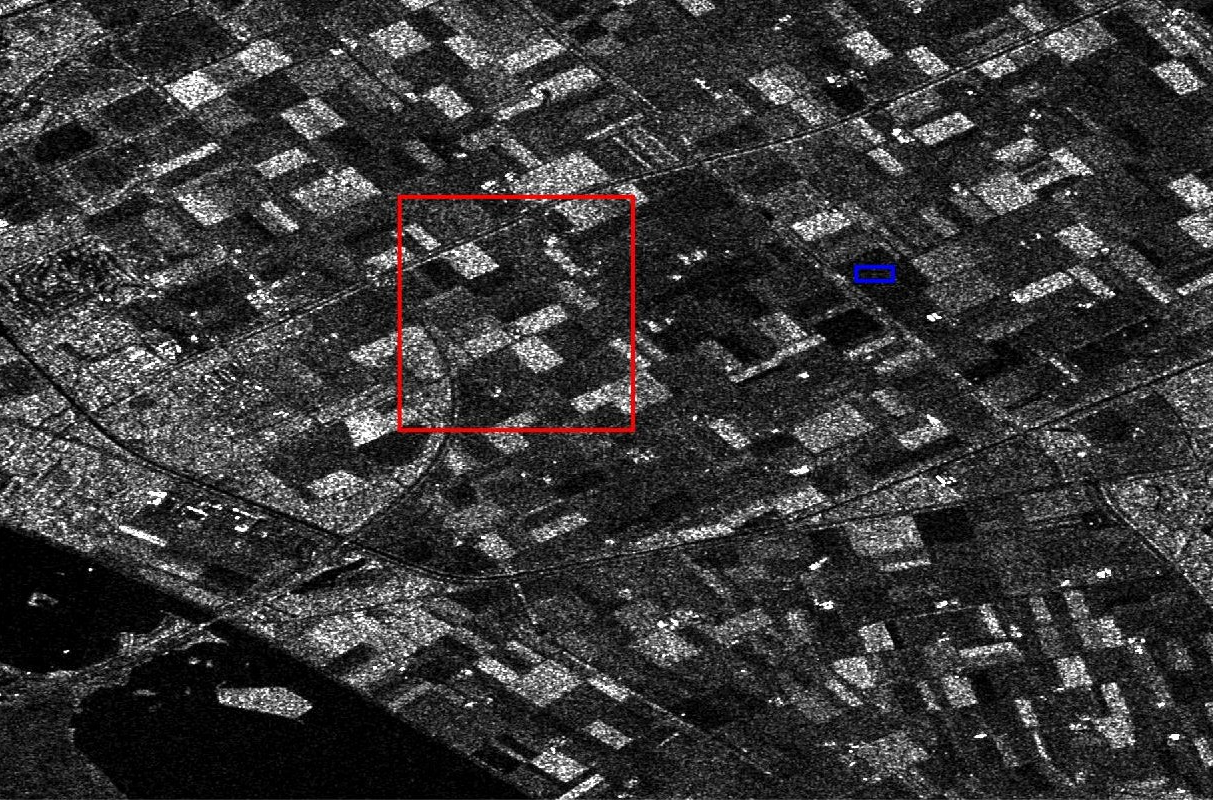} &
		\includegraphics[width=5cm,height=4cm]{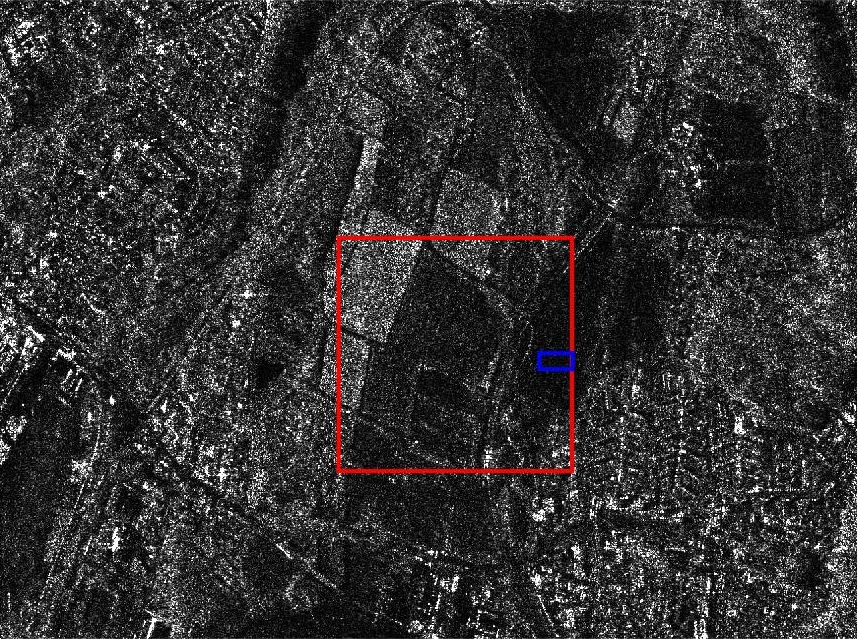} \\
		{(d)}&{(e)} &{(f)}\\
	\end{tabular}
	\caption{The input images used to test deSpeckNet. The area in the red and green boxes were used for qualitative comparison of the methods, whereas the area in blue was used to estimate quality metrics such as equivalent number of looks (ENL) and coefficient of variation (Cx). (a) Indonesia, (b) DRC, (c) The Netherlands-Utrecht, (d) Japan,  (e) The Netherlands-Flevoland, (f) Germany. All the test images used multi-looked images except the Iceye image in Germany which was a single look image. }\label{fig:Fig55}
\end{figure*}

\section{Experimental Setup}
\subsection{Label}

\noindent We use the reference label preparation method suggested in \cite{chierchia2017sar} that uses a large stack of multi-temporal images to create the label clean image. Since this reference label is prepared from real SAR images, it represents the properties of real SAR features. To achieve the best results, the patches selected for training have to be stable in time, \emph{i.e.} the scene must not have large temporal variation. This can be ensured by using the standard deviation of intensity for each pixel in the image as :

\begin{equation}
  Z = \sqrt{\frac{\sum_{i=1}^N (x_i - \tilde{x})^2}{N-1}}. 
\end{equation}

\noindent Here $x$ is the pixel intensity, $\tilde{x}$ is the mean of the temporal pixel intensity value, $N$ is the number of images in the temporal stack. If the standard deviation of pixel intensity values over the period is above a threshold ($\nu = 0.1$), that pixel in the temporally averaged image is masked. In places where we have low standard deviation below a pre-defined threshold we can apply a temporal averaging on the image as:\\

\begin{equation}
 \mathrm{x} =  \left\{
                \begin{array}{ll}
                  \frac{1}{N} \sum_{i=1}^N x_i ~~ \mathrm{if}~~Z \leq 0.1\\
                  0 ~~\mathrm{otherwise}
                \end{array}.
\right. 
\end{equation}

\noindent In this way, we can prepare a temporal average of patches to be used as reference labels, whose number is detailed in Table I. {In cases where the number of available multi-temporal images is limited, a spatial multilooking is recommended to suppress residual speckle noise in the synthesized reference label image \cite{zhao2019ratio}. In our experiments, we have used 23 to 28 multi-temporal Sentinel-1 GRD images that were originally multilooked 5 times, so we didn't perform additional multilooking to synthesize the reference label images. } \\

\subsection{Training}

\noindent To investigate the performance of deSpeckNet, we trained the network in a temporally stationary scene (S1-Indonesia) and to investigate the tuning capability of the designed architecture we applied it to an image acquired in different geographic regions and landcover types and sensors.  In both $\mathrm{FCN_{clean}}$ and $\mathrm{FCN_{noise}}$, we  used 17 blocks consisting of convolution, batch normalization and non-linear activation, determined empirically. The details of each block is shown in Table~\ref{tbl:Tab3}. \\

\diego{\noindent On the $\mathrm{FCN_{clean}}$ side, we used {the same} architecture as that of the noise side. We used a $L_{MSE}$ and a combination of total variation $L_{TV}$ and $L_{MSE}$ loss function that uses the temporally averaged SAR images as reference label. When tuning the network to a new region or data type, we used only the Total Variation $L_{TV}$ loss function that relied on the already learned Gamma distribution of the speckle noise. Hence, {we did not} need to use any reference label to tune the network. In the element wise multiplication the original image is reconstructed and an $L_{MSE}$ loss function is implemented that used the original noisy image as a reference label. }{} \\

\begin{table}[tbh]
\begin{center}
\small
\setlength\tabcolsep{6pt}
\begin{tabular}{lccc} \hline
 Layer & Module type & Dimension \\ \hline
Conv1& $\mathrm{Conv}$ & 3 $\times$ 3$\times$ 1 $\times$ 64  \\ 
 & $\mathrm{ReLU}$ &  \\
Conv2 ($\times 15$)& $\mathrm{Conv}$ & 3 $\times$ 3$\times$ 64 $\times$ 64 \\ 
& $\mathrm{BN}$ &  \\
 & $\mathrm{ReLU}$ &  \\
 \\
 Prediction& $\mathrm{Conv}$ & 3 $\times$ 3$\times$ 64 $\times$ 1 \\ \hline
\end{tabular}
\end{center}
\caption{Architecture of the FCN blocks of deSpeckNet.}\label{tbl:Tab3}
\end{table}

\noindent To train deSpeckNet in the initial phase, we prepared the input noisy images and their corresponding reference labels into a $40\times 40$ patches. We created an overall 117,888 patches for training. {Since, deSpeckNet is designed to fine tune images with different noise statistics than what it was trained on, there is no need for a training set that represents well the test set, and thus, the diversity of the training set becomes less of a limiting factor.} We used a batch size of 128 so at every epoch the network used 921 iterations. We trained the network for a total of 30 epochs using a learning rate between  $10^{-3}$ and $10^{-4}$ by decreasing the learning rate by 0.002 every 10 epochs.  In the initial training phase we set the weight ($\lambda$) of the $L_{TV}$ to zero and the $L_{Clean}$ was given a $\mu$  of 1 and $\xi$ of $L_{Noisy}$ was set to 0.01.  To fine-tune the network for new regions or new set of data we used the same learning rate as the initial training phase for one epoch.  In the fine tuning phase, for the Sentinel-1 GRD images we set the $\lambda$ of the $L_{TV}$ to $0$ and the $L_{Clean}$ was given a $\mu$ of $10^{-2}$ and $\xi$ of $L_{Noisy}$ was set to 1. For single look datasets such as the Iceye X2 images we set the $\lambda$ of the $L_{TV}$ to $10^{-4}$ and the $L_{Clean}$ was given a $\mu$ of $10^{-2}$ and $\xi$ of $L_{Noisy}$ was set to 1. In both cases we used an Adam optimizer \cite{kingma2014adam} whose decay rate was fixed at $0.9$. We used early stopping in fine tuning the model to reconstruct the new images without reference labels. {To properly evaluate the performance of the network we trained the network 10 times from random seeds using the improved Xavier weight initialization \cite{glorot2010understanding}}. \\

\begin{figure*}
	\centering
	\begin{tabular}{cccccc}
		\includegraphics[width=0.15\textwidth]{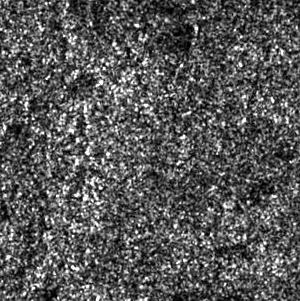}&
		\includegraphics[width=0.15\textwidth]{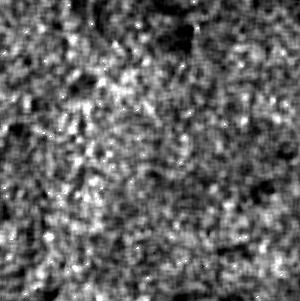}&
		\includegraphics[width=0.15\textwidth]{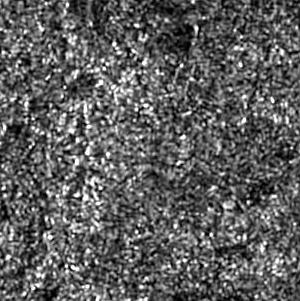}&
		\includegraphics[width=0.15\textwidth]{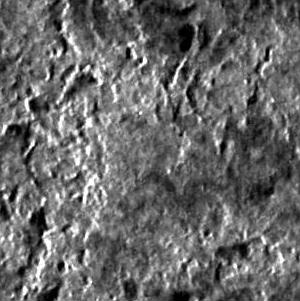}&
		\includegraphics[width=0.15\textwidth]{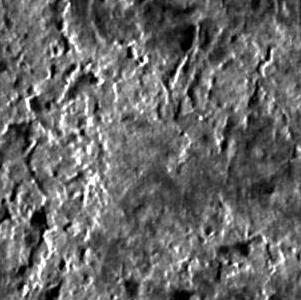}&
		\includegraphics[width=0.15\textwidth]{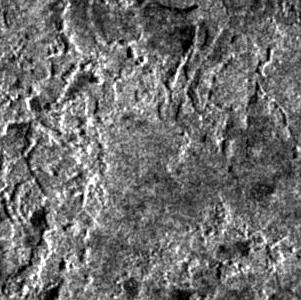}\\
		
	\end{tabular}
	
		\begin{tabular}{cccccc}
		\includegraphics[width=0.15\textwidth]{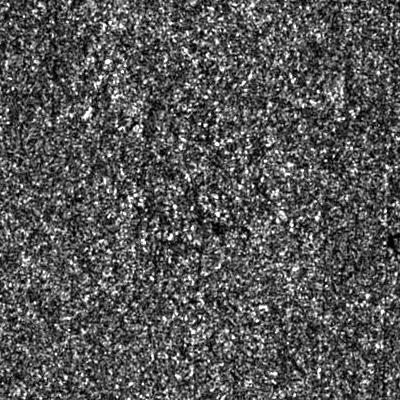}&
		\includegraphics[width=0.15\textwidth]{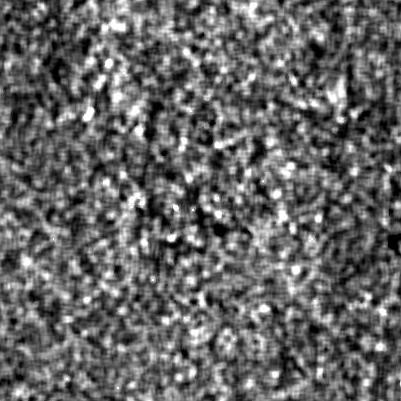}&
		\includegraphics[width=0.15\textwidth]{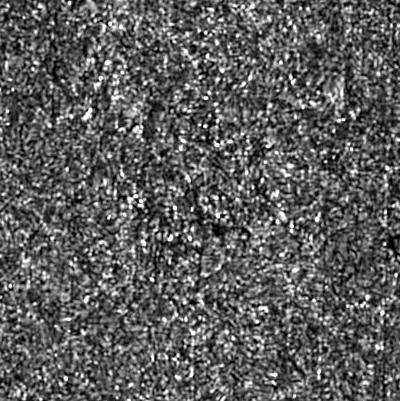}&
		\includegraphics[width=0.15\textwidth]{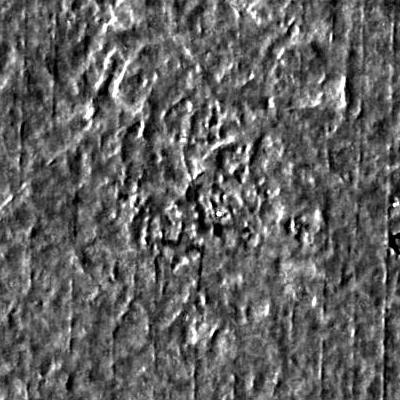}&
		\includegraphics[width=0.15\textwidth]{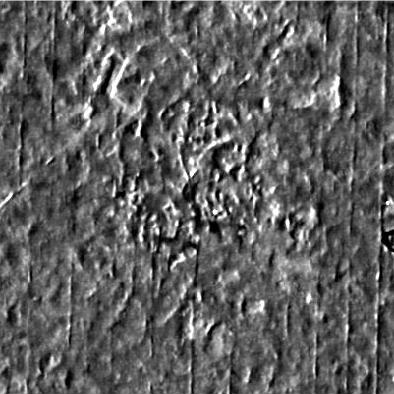}&
		\includegraphics[width=0.15\textwidth]{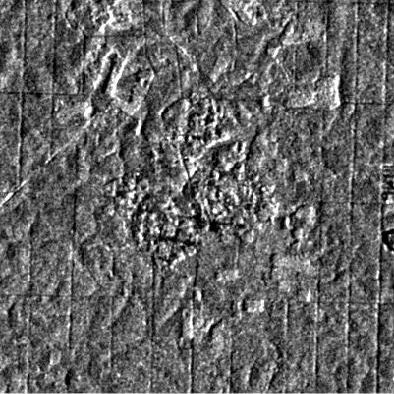}\\
	
	{(a)}&{(b)}& {(c)} &{(d)} &{(e)} &{(f)}
		
	\end{tabular}

	\caption{Despeckling result for different baseline methods and deSpeckNet in the Indonesia Sentinel-1 image. We show a $300 \times 300$ and $400 \times 400$ image patch for (a) input noisy image, {(b) Lee-sigma}, (c) SAR-BM3D, (d) SAR-CNN, (e) deSpeckNet. (f) is the temporal average image used to estimate PSNR/SSIM/DG. {Some residual noise can be observed in the temporal average image because we averaged 23 images. However, due to the MSE loss function, its effect on the training performance is negligible.}  }\label{fig:Fig3}
\end{figure*}

\begin{figure*}
	\centering
	\begin{tabular}{cccccc}
		\includegraphics[width=0.15\textwidth]{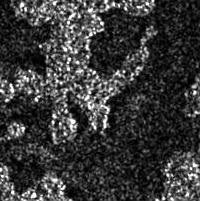}&
		\includegraphics[width=0.15\textwidth]{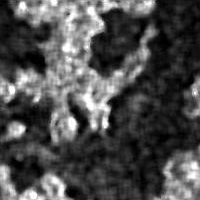}&
		\includegraphics[width=0.15\textwidth]{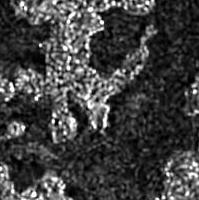}&
		\includegraphics[width=0.15\textwidth]{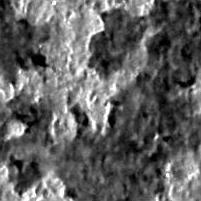}&
		\includegraphics[width=0.15\textwidth]{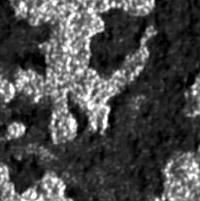}&
		\includegraphics[width=0.15\textwidth]{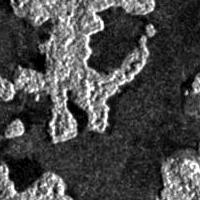}\\
	{(a)}&{(b)}& {(c)} &{(d)} &{(e)} &{(f)}
	\end{tabular}
	
	\caption{Despeckled images in the DRC Sentinel-1 image. We show a $200 \times 200$ image patch for (a) input noisy image, {(b) Lee-sigma}, (c) SAR-BM3D, (d) SAR-CNN, (e) deSpeckNet. (f) is the temporal average image used to estimate PSNR/SSIM/DG. This area is selected to demonstrate the generalization capability of the model in a similar landcover types from what it was trained on but different geographic region. }\label{fig:Fig4}
\end{figure*}

\noindent To train deSpeckNet, we used the MatConvNet {framework} \cite{vedaldi2015matconvnet} in a Matlab 2018a environment run on a Linux operating system with Intel® Xeon(R) E-2176M CPU and Quadro P2000 GPU. \\

\begin{figure*}
	\centering
	\begin{tabular}{cccccc}
		\includegraphics[width=0.15\textwidth]{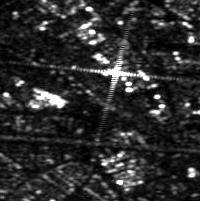}&
		\includegraphics[width=0.15\textwidth]{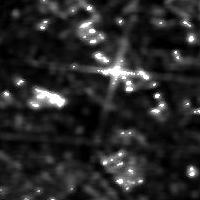}&
		\includegraphics[width=0.15\textwidth]{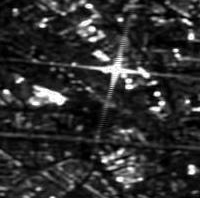}&
		\includegraphics[width=0.15\textwidth]{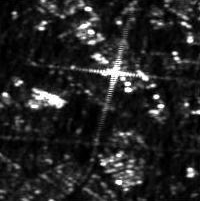}&
		\includegraphics[width=0.15\textwidth]{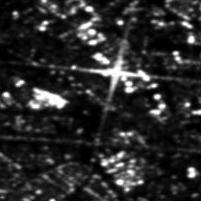}&
		\includegraphics[width=0.15\textwidth]{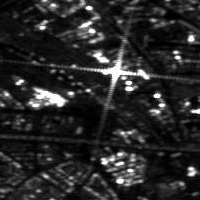}\\
		{(a)}&{(b)}& {(c)} &{(d)} &{(e)} &{(f)}
	\end{tabular}

	\caption{Despeckled images in the Netherlands-Utrecht Sentinel-1 image. We show a $200 \times 200$ tile for (a) input noisy image, {(b) Lee-sigma}, (c) SAR-BM3D, (d) SAR-CNN, (e) deSpeckNet. (f) is the temporal average image used to estimate PSNR/SSIM/DG. This area is selected to demonstrate the scale-ability of the model in a similar sensor types from what it was trained on but different geographic region and landcover type. The temporal average image displayed is used only for deriving the quality metrics. ( {In Figure 6f., for visualization purposes we didn't mask pixels that did not fulfill the temporal standard deviation criteria}).  }  \label{fig:Fig7}
\end{figure*}

\subsection{Quality metrics}

\noindent In test areas where we have a multi-temporal images and where the assumption of temporal stationarity was fulfilled, we used the temporally averaged images as validation ground truth images to derive quality metrics. {The quality metrics used to evaluate the performance of deSpeckNet in the presence of full reference label image are the peak signal to noise ratio (PSNR), structural similarity index (SSIM), despeckling gain (DG), edge preservation index (EPI) and the strong scatterer preservation index ($\mathrm{C}_{NN}$).} PSNR estimates the quality of the reconstructed noise free image resemblance to the reference data, in this case the temporally averaged Sentinel-1 image as:

\begin{equation}
    PSNR = 20~ \mathrm{log}_{10} (\frac{\hat{X}_{max}}{\sqrt{MSE}}).
\end{equation}

\noindent Here, $\hat{X}_{max}$ is the maximum power given as 255. We do this by converting the 32bit data to 8bit data. The MSE is computed between the label (${X}$) and the reconstructed images ($\hat{X}$) given as $\mathrm{MSE} = \mathrm{E}[(\hat{X}-X)^2]$.\\

\noindent SSIM estimates the structural similarity between the label (${X}$) and the reconstructed image ($\hat{X}$) as:

\begin{equation}
    SSIM(\hat{X}, {X}) = \frac{(2 \mu_{\hat{X}} \mu_{{X}} + c_1) (2 \sigma_{\hat{X} {X}} + c_2)}{(\mu_{\hat{X}}^2 + \mu_{{X}}^2 + c_1) (\sigma _{\hat{X}}^2 + \sigma _{{X}}^2 + c_2 )},
\end{equation}

\noindent Where  $\mu_{X}$ is the mean of image $X$ and $\sigma_X$ is its standard deviation.\\

\noindent {The despeckling gain (DG) estimates the speckle rejection ability of a particular despeckling method \cite{di2013benchmarking}. Therefore, a large DG value indicates a higher speckle removal ability. DG is estimated as follows: 
}

\begin{equation}
    DG = 10 \log_{10}(\frac{\mathrm{MSE}(\hat{X},Y)} {\mathrm{MSE}(\hat{X}, X)}).
\end{equation}

\noindent {To evaluate edge preservation, we use the edge preservation index (EPI) \cite{mullissa2017scattering} \cite{foucher2014analysis}. EPI is derived by first defining the gradient preservation index (GP){, which is} the ratio between the gradient values in the filtered intensity image ($\hat{X}$) and the gradient of the reference image (${X}$). } \\

\begin{equation}
    {GP = \frac{\sum \nabla_{\hat{X}}}{\sum \nabla_{X}}}.
\end{equation}

\noindent {Here, $\nabla$ is the Sobel gradient operator. {EPI is calculated} by projecting the GP values in the interval $[0,1]$ using a triangular equation as:} \\

\begin{equation}
 {EPI =  \left\{
                \begin{array}{ll}
                  1-|1-GP| ~~ \mathrm{if}~~GP < 2\\
                  0 ~~\mathrm{otherwise}
                \end{array}.
\right.} 
\end{equation}

\noindent {The strong scatterer preservation index ($\mathrm{C}_{NN}$) estimates the strong scatterer preservation ability of a particular filter \cite{di2013benchmarking}. Similarly, a higher $\mathrm{C}_{NN}$ indicates a higher preservation of a strong scatterer. $\mathrm{C}_{NN}$ is given as:}

\begin{equation}
    C_{NN} = 10 \log_{10} \frac{X_{CF}}{X_{NN}},  
\end{equation}

\noindent {where $X_{CF}$ is the intensity observed at the strong scatterer site and $X_{NN}$ is  the average intensity of the surrounding eight pixels.}\\

\noindent In test regions where a temporal average image is not available, as a metric for quality of performance, we used visual inspection as a qualitative measure and the equivalent number of looks (ENL) and the coefficient of variation ($\mathrm{C}_x$) derived in a homogeneous region as a quantitative measure for comparison. The ENL is derived by taking the ratio between the mean square ($\mu^2$) and variance ($\sigma^2$) of a homogeneous region in the image as:

\begin{equation}\label{eq:13}
    ENL = \frac{\mu_{\hat{X}}^2}{\sigma_{\hat{X}}^2}.
\end{equation}

\noindent {Whereas $\mathrm{C}_x$ is a measure for the preservation of texture in the filtered image and is derived in a homogeneous region by taking the the ratio of the standard deviation with the mean intensity, given as}:

\begin{equation}\label{eq:26}
    C_x = \frac{\sigma_{\hat{X}}}{\mu_{\hat{X}}}.
\end{equation}

\noindent As can be seen from (\ref{eq:13}) in uniform regions the filter that achieves the smallest variance in pixel intensities i.e. best despeckling performance, will result in the highest ENL and lowest $C_x$ values. \\ 

\begin{figure*}
	\centering
	\begin{tabular}{cccccc}
		\includegraphics[width=0.17\textwidth]{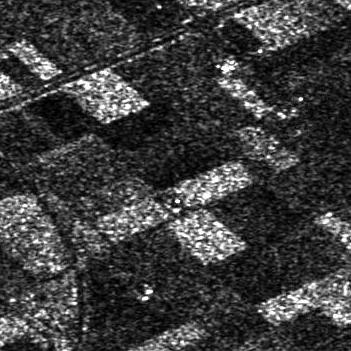}&
		\includegraphics[width=0.17\textwidth]{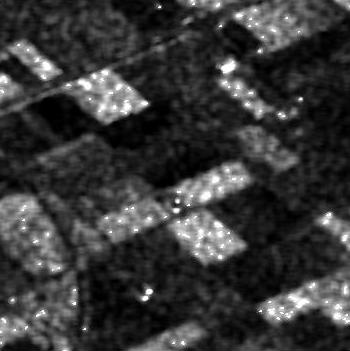}&
		\includegraphics[width=0.17\textwidth]{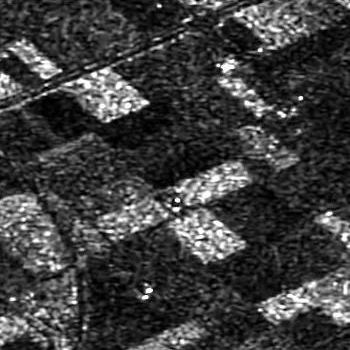}&
		\includegraphics[width=0.17\textwidth]{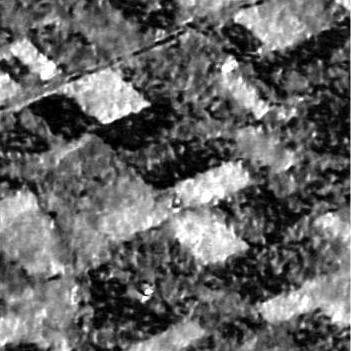}&
		\includegraphics[width=0.17\textwidth]{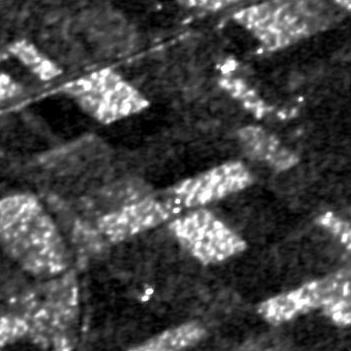}\\
		{(a)}&{(b)}& {(c)} &{(d)} &{(e)} 
	\end{tabular}
	\caption{Despeckled images in the Netherlands-Flevoland Sentinel-1 image. We show a $350 \times 350$ image patch for (a) input noisy image, {(b) Lee-sigma}, (c) SAR-BM3D, (d) SAR-CNN, (e) deSpeckNet. This area is selected to demonstrate the generalization capability of the model in a similar sensor type from what it was trained on but different landcover type.}\label{fig:Fig8}
\end{figure*}

\section{Results and Discussion} 

\noindent To assess the performance of deSpeckNet, we compared it with { the improved Lee-sigma filter \cite{lee2008improved},}  SAR-BM3D~\cite{dabov2007image}, {an unsupervised method based on block-matching}, and SAR-CNN \cite{chierchia2017sar}, {a supervised CNN-based method,} both qualitatively and quantitatively. We selected SAR-CNN because it is designed to be trained on real SAR images as opposed to the methods trained on simulated noise and optical images. \\

\subsection{Evaluation on the same region used for training}

\noindent In the Indonesia image, {a forested landscape for which a temporal average is available as clean reference}, { improved Lee sigma filter failed to remove speckle and {to} preserve the subtle features in the image}. SAR-BM3D, {being an unsupervised method that does not make use of the reference image and benefits from highly structured elements in the image to perform non-local matching,} performed sub-optimally in preserving edges and removing noise from homogeneous regions {due to the lack of strong structures}. {deSpeckNet and SAR-CNN performed similarly, since they are both trained with supervision on this image. Both are better than SAR-BM3D} in preserving features and removing noise from homogeneous regions (Figure~\ref{fig:Fig3}, Table~\ref{tbl:Tab4}).  \\

\begin{table*}[tbh]
	\begin{center}
		\small
		\setlength\tabcolsep{12pt}
		\begin{tabular}{lccccc|c} \hline
		Test area & Metric & {Lee Sigma} & SAR-BM3D & SAR-CNN & deSpeckNet & SAR-CNN (*test)  \\ \hline
Indonesia & PSNR &  {39.10} &  36.14 & 45.67 $\pm$ 0.07 & \textbf{45.70} $\pm$ 0.05  & 45.67 $\pm$ 0.07\\
		  & SSIM &  {0.93} &  0.86 & 0.97 $\pm$ 0.00 & \textbf{0.97} $\pm$ 0.00  & 0.97 $\pm$ 0.00\\
		  & DG &  {4.76} &  1.77 & 11.49 $\pm$ 0.07 & \textbf{11.51} $\pm$ 0.05  & 11.49 $\pm$ 0.07\\
		  & EPI &  {0.20}&  0.86 & 0.90 $\pm$ 0.02 & \textbf{0.92} $\pm$ 0.01 & 0.90 $\pm$ 0.02\\
		  & ENL &  {30.51}&  11.67 & 105.35 $\pm$ 4.53 & \textbf{114.51} $\pm$ 7.12 & 105.35 $\pm$ 4.53\\ 
		  & Cx &  {0.18} &  0.29 & \textbf{0.09} $\pm$ 0.002 & {0.09} $\pm$ 0.003  & 0.09 $\pm$ 0.002\\\hline
DRC       & PSNR  & {37.34}&  35.52 & 38.45 $\pm$ 0.00 & \textbf{39.33} $\pm$ 0.01 &{39.91} $\pm$ 0.01 \\ 
	      & SSIM  & {0.89} &0.85 & 0.90 $\pm$ 0.00 & \textbf{0.92} $\pm$ 0.00 &{0.92} $\pm$ 0.00\\ 
	      & DG &  {3.3} &  1.47 & 4.45 $\pm$ 0.06 & \textbf{5.3} $\pm$ 0.008  & 5.82 $\pm$ 0.04\\
		  & EPI  & {0.31} & 0.84 & 0.74 $\pm$ 0.01 & \textbf{0.92} $\pm$ 0.002 &{0.91} $\pm$ 0.01\\
		  & ENL  & {41.53}& 11.04 & 27.72 $\pm$ 5.26 & \textbf{118.21} $\pm$ 17.99 &150.29 $\pm$ 38.93\\
		  & Cx &  {0.18} &  0.30 & 0.18 $\pm$ 0.026 & \textbf{0.09} $\pm$ 0.008  & 0.07 $\pm$ 0.01\\\hline
NL-Utrecht & PSNR  & {26.89}& {28.02} & 19.88 $\pm$ 1.75 & \textbf{28.14} $\pm$ 0.32 &  {29.97} $\pm$ 0.50\\ 
		   & SSIM  & {0.77}& 0.81 & 0.66 $\pm$ 0.00 & \textbf{0.84} $\pm$ 0.00 & {0.85} $\pm$ 0.00 \\
		   & DG &  {-0.3} &  0.82 & -7.25 $\pm$1.87 & \textbf{0.94} $\pm$ 0.32  & 1.72 $\pm$ 0.52\\
		   & EPI  & {0.53} & \textbf{0.97} & 0.88 $\pm$ 0.01 & {0.87} $\pm$ 0.03& {0.97} $\pm$ 0.02 \\
		   & ENL  & {100.66}& 80.47 & 57.36 $\pm$ 20.69 & \textbf{270.20} $\pm$ 98.90& {454.54} $\pm$ 139.52\\ 
		   & Cx &  \textbf{0.009} &  0.15 & 0.13 $\pm$ 0.017 & {0.062} $\pm$ 0.0087  & 0.04 $\pm$ 0.007\\
		   & $C_{NN}$ &  \textbf{2.84} &  2.57 & 2.83 $\pm$ 0.41 & {2.08} $\pm$ 0.14  & 2.84 $\pm$ 0.08\\\hline
Japan      & ENL  & {60.83}&42.45 & 8.66 $\pm$ 0.21 & \textbf{114.95} $\pm$ 25.39&- \\  
           & Cx &  {0.11} &  0.10 & 0.31 $\pm$ 0.008 & \textbf{0.09} $\pm$ 0.004  & -\\\hline
NL-Flevoland & ENL  & {41.40}& 12.12 & 5.56 $\pm$ 2.59 & \textbf{132.37} $\pm$ 40.05 &- \\ 
            & Cx &  {0.15} &  0.28 & 0.44 $\pm$ 0.07 & \textbf{0.08} $\pm$ 0.008  & -\\\hline
Germany    & ENL  & {29.36}& 5.69 & 3.43 $\pm$ 0.13 & \textbf{65.86} $\pm$ 1.18&- \\ 
           & Cx &  {0.18} &  0.41 & 0.53 $\pm$ 0.00 & \textbf{0.12} $\pm$ 0.00  & -\\\hline
			
		\end{tabular}\\
	\end{center}
	\caption{Quantitative quality metrics for all the test images. {SAR-CNN (*test) refers to SAR-CNN  fine tuned on a temporally averaged image for the target scene and it is used} as an upper bound. {All results report an average over 10 runs and the corresponding standard deviation. The Lee sigma and SAR-BM3D are not initialized randomly, hence, the uncertainties are not shown.}}\label{tbl:Tab4}
\end{table*}

\begin{figure*}
	\centering
	\begin{tabular}{ccccc}
		\includegraphics[width=0.17\textwidth]{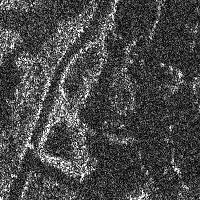}&
		\includegraphics[width=0.17\textwidth]{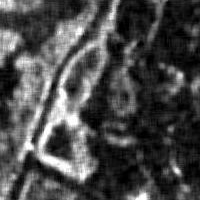}&
		\includegraphics[width=0.17\textwidth]{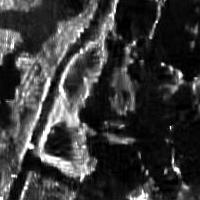}&
		\includegraphics[width=0.17\textwidth]{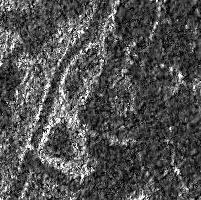}&
		\includegraphics[width=0.17\textwidth]{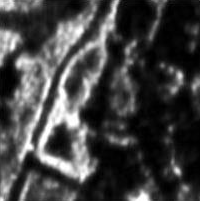}\\

	\end{tabular}
		\begin{tabular}{ccccc}
		\includegraphics[width=0.17\textwidth]{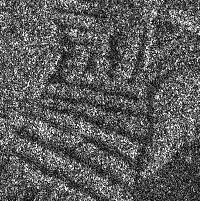}&
		\includegraphics[width=0.17\textwidth]{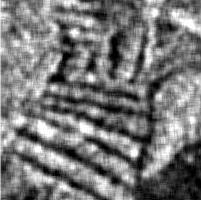}&
		\includegraphics[width=0.17\textwidth]{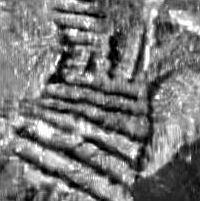}&
		\includegraphics[width=0.17\textwidth]{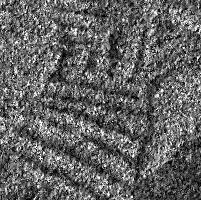}&
		\includegraphics[width=0.17\textwidth]{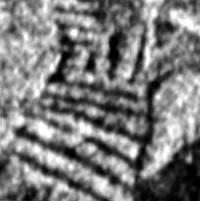}\\
		{(a)}&{(b)}& {(c)} &{(d)} &{(e)}
	\end{tabular}
	\caption{Despeckled images in the Japan ALOS2-PALSAR2 image. We show two $200 \times 200$ image patches for (a) input noisy image, {(b) Lee-sigma}, (c) SAR-BM3D, (d) SAR-CNN, (e) deSpeckNet. This area is selected to demonstrate the generalization capability of the model in a similar landcover types from what it was trained on but different sensor. }\label{fig:Fig6}
\end{figure*}

\subsection{Tests on other Sentinel scenes}
\subsubsection{Qualitative results}

\noindent {On the remaining images no reference clean image is used to train SAR-CNN nor deSpeckNet. In the case of SAR-CNN, we apply the model that was trained on the Indonesia image (Two image subset are presented in (Figure~\ref{fig:Fig3})). {The Lee sigma filter and} SAR-BM3D, being unsupervised, work on the same setting as in the Indonesia image.} 
{On the DRC image (Figure~\ref{fig:Fig4}), which is similar in nature to the one over Indonesia}, {both the improved Lee sigma,} SAR-BM3D still perform sub-optimally in preserving features and removing noise from homogeneous regions or preserving subtle features. SAR-CNN filtered the DRC image {using the noise model learned from the Indonesia image. Although both are obtained with Sentinel-1, the differences in the noise distribution are large enough to} result in blurred edges and unevenly filtered noise from homogeneous regions. deSpeckNet, however, was able to be tuned to the DRC image without using any new clean reference labels (Figure~\ref{fig:Fig4}e), removing much of the noise from homogeneous regions while preserving the edges between regions. \\

\begin{figure*}
	\centering
	\begin{tabular}{cccccc}
		\includegraphics[width=0.17\textwidth]{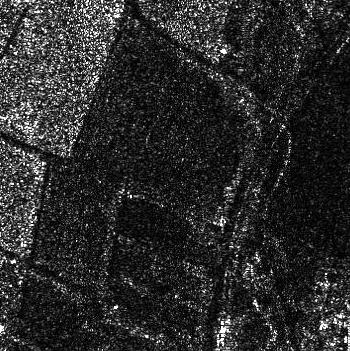}&
		\includegraphics[width=0.17\textwidth]{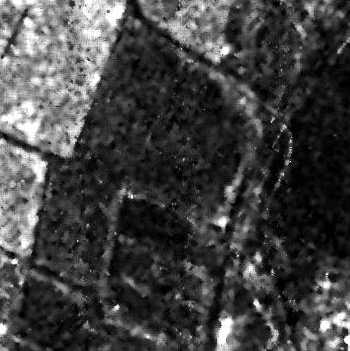}&
		\includegraphics[width=0.17\textwidth]{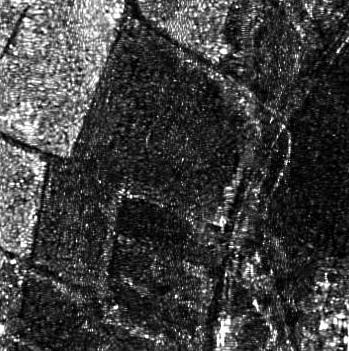}&
		\includegraphics[width=0.17\textwidth]{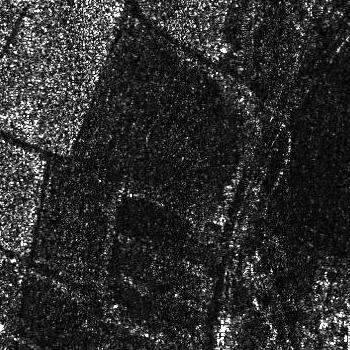}&
		\includegraphics[width=0.17\textwidth]{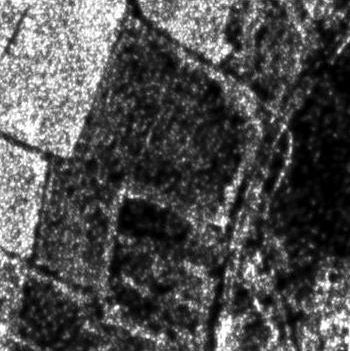}\\
		{(a)}&{(b)}& {(c)} &{(d)} &{(e)}
	\end{tabular}
	\caption{Despeckled images in the Kiel Iceye x-2 image. We show a $350 \times 350$ image patch for (a) input noisy image, {(b) Lee-sigma}, (c) SAR-BM3D, (d) SAR-CNN, (e) deSpeckNet. This area and data type is selected to demonstrate the generalization capability of the model in a different sensor types, geographic region and landcover type. }\label{fig:Fig9}
\end{figure*}

\noindent In the Netherlands-Utrecht image (Figure~\ref{fig:Fig7}), {the improved Lee sigma filter over-filtered and distorted all features in the image.} Whereas,  SAR-BM3D performed better than in the previous images as the image contrast was higher than the Indonesia and DRC case. However, it suffered from over-filtering, resulting in the smoothing out of subtle features. SAR-CNN resulted in sub-optimal results as it failed to adequately remove the noise from homogeneous regions and improve the overall signal to noise ratio of the image. deSpeckNet {succeeded} in preserving features and adequately removing noise from homogeneous regions (Figure~\ref{fig:Fig7}d) when compared to SAR-CNN. {The advantage} of deSpeckNet was {also} demonstrated when applied to the image in the Netherlands-Flevoland. Here, {the improved Lee sigma filter overfiltered the scene resulting in blurred features and} SAR-BM3D was not able to remove the speckle maintaining the noisy appearance, whereas SAR-CNN indiscriminately filtered the image distorting many of the image features. On the contrary, deSpeckNet removed the speckle from homogeneous regions while maintaining subtle features in the image (Figure~\ref{fig:Fig8}e).  \\

\subsubsection{Quantitative results}
\noindent The capability of deSpeckNet is further exemplified by the improvement of the quantitative metrics of PSNR, SSIM, DG, {EPI} and ENL when compared with the {improved Lee sigma}, SAR-BM3D and SAR-CNN. {The ENL is estimated in a  minimum window size of $25\times 55$ pixel window for the Flevoland test area and a maximum of $80\times 115$ pixel window for the Japan test area Figure~\ref{fig:Fig55}.} In the Indonesia and DRC test areas deSpeckNet achieved the highest PSNR, SSIM, DG, {EPI} and ENL values than SAR-BM3D and SAR-CNN (Table~\ref{tbl:Tab4}).  This trend was slightly changed when comparing the quantitative results from the Netherlands-Utrecht image. Here, SAR-BM3D achieved a slightly higher {EPI} { and $C_{NN}$} when compared with deSpeckNet. This is due to the fact that SAR-BM3D {is better adapted to highly structured images {with distinct strong scatterers}, such as this urban scene. However, deSpeckNet results in a higher {PSNR}, SSIM and DG value than SAR-BM3D, which suggests that the latter might be incurring in {inconsistencies} in filtering, to which the localized EPI {and $C_{NN}$} metric is less sensitive.} SAR-CNN achieved overall low values in  all compared metrics {due to its inability to adapt to new noise distributions}.  A similar trend was also observed in the Netherlands-Flevoland image (Table~\ref{tbl:Tab4}), where deSpeckNet achieved the significantly higher ENL values and the smallest $C_x$ than all the other baseline methods. {The improved Lee sigma filter achieved the lowest PSNR, SSIM, DG,  EPI and ENL values in all test images showing a sharp contrast between the traditional localized speckle filters and machine learning based filters.} To establish the upper bound for tuning, we did a supervised tuning of SAR-CNN in the DRC {and Netherlands-Utrecht} image by using the temporally averaged image. The SAR-CNN supervised  fine tuning achieved a mean PSNR of 39.91, SSIM of 0.92 {and DG of 5.82} {in the DRC and a mean PSNR of 29.97, SSIM of 0.85 {and DG of 1.72} in the Netherlands-Utrecht image (Table~\ref{tbl:Tab4})}, which was slightly higher than the values achieved by deSpeckNet. Note that these results show that deSpeckNet, ever without using supervised tuning on the test image, is able to reach an equivalent performance of a network tuned in a clean test image, which, the more often, is not available. 

\begin{figure*}
	\centering
	\begin{tabular}{ccc}
		\includegraphics[width=5cm,height=4cm]{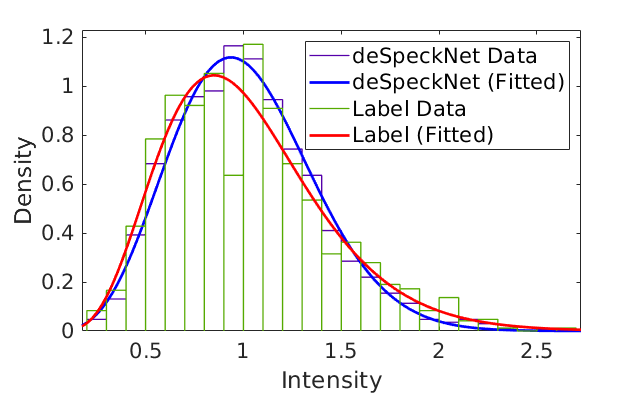}&
		\includegraphics[width=5cm,height=4cm]{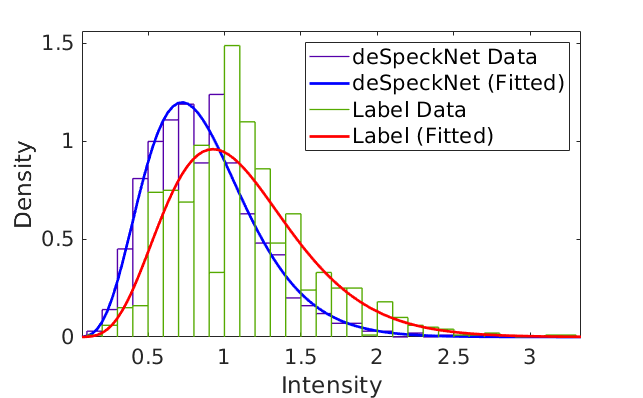} &
		\includegraphics[width=5cm,height=4cm]{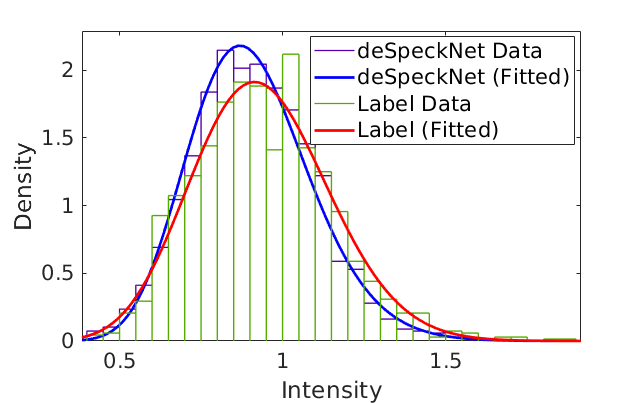} \\
		{(a)}&{(b)} &{(c)}
	\end{tabular}
	\begin{tabular}{ccc}
		\includegraphics[width=5cm,height=4cm]{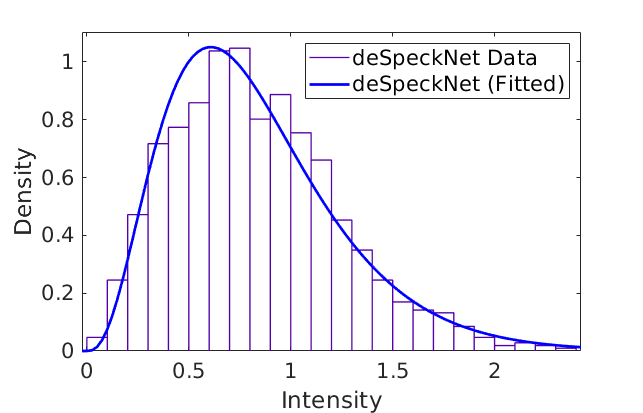}&
		\includegraphics[width=5cm,height=4cm]{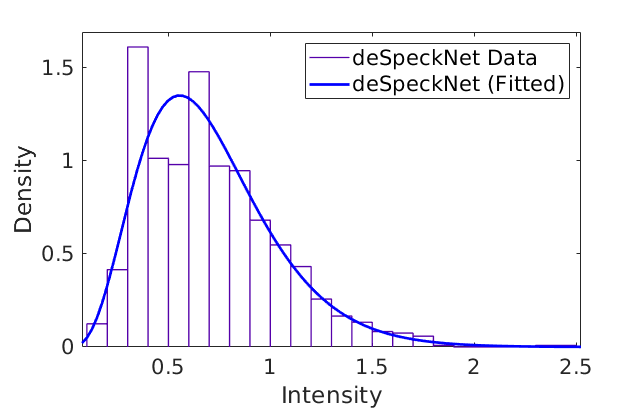} &
		\includegraphics[width=5cm,height=4cm]{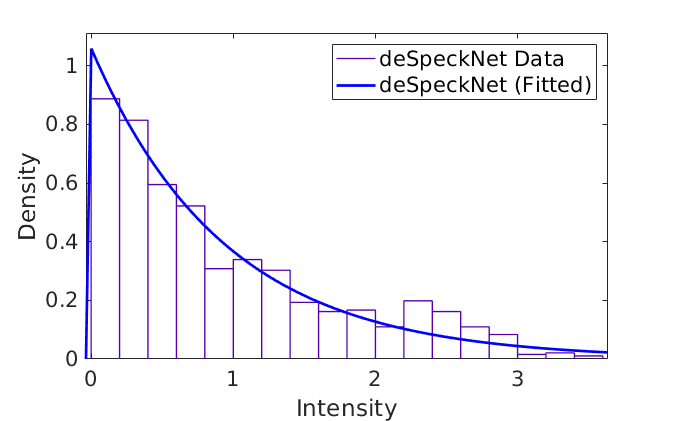} \\
		{(d)}&{(e)} &{(f)}\\
	\end{tabular}
	\caption{The noise distribution estimated by deSpeckNet and the speckle noise derived from the reference data for each of the test areas. (a) , (b) DRC, (c) The Netherlands-Utrecht, (d) Japan,  (e) The Netherlands-Flevoland, (f) Germany. All the test areas were from Multi-looked images except the Iceye image in Germany which was a single look image. }\label{fig:Fig5}
\end{figure*}

\begin{figure*}
	\centering

\begin{tabular}{cccccc}
	\includegraphics[width=0.15\textwidth]{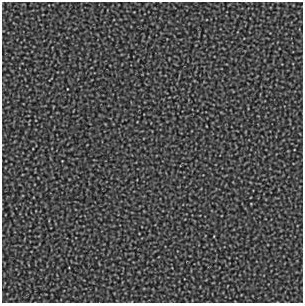}&
	\includegraphics[width=0.15\textwidth]{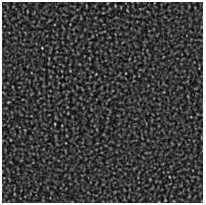}&
	\includegraphics[width=0.15\textwidth]{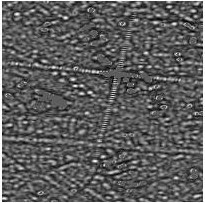}&
	\includegraphics[width=0.15\textwidth]{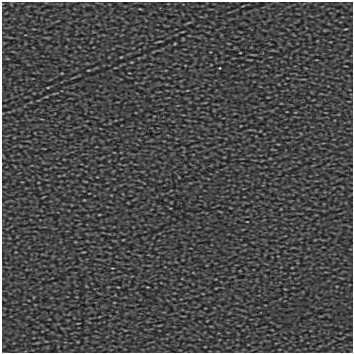}&
	\includegraphics[width=0.15\textwidth]{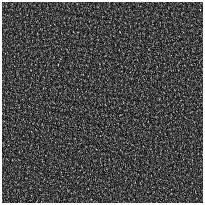}&
	\includegraphics[width=0.15\textwidth]{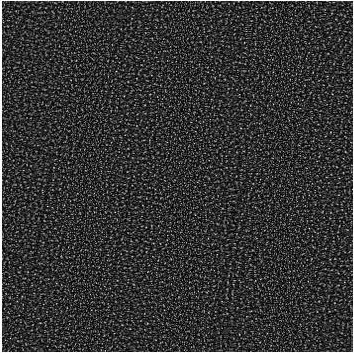}\\
	
\end{tabular}

\begin{tabular}{cccccc}
	\includegraphics[width=0.15\textwidth]{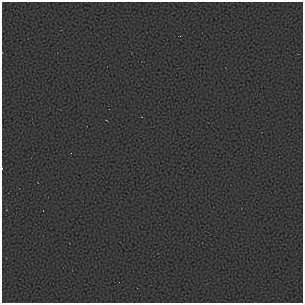}&
	\includegraphics[width=0.15\textwidth]{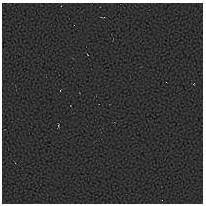}&
	\includegraphics[width=0.15\textwidth]{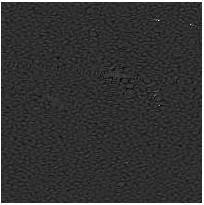}&
	\includegraphics[width=0.15\textwidth]{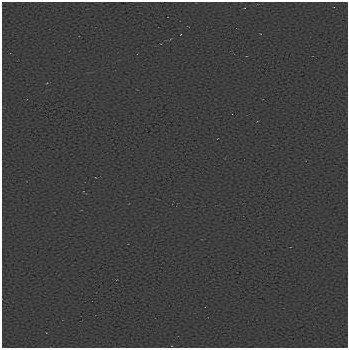}&
	\includegraphics[width=0.15\textwidth]{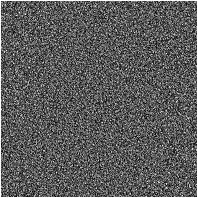}&
	\includegraphics[width=0.15\textwidth]{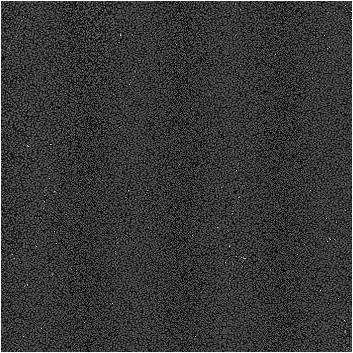}\\
	
\end{tabular}

\begin{tabular}{cccccc}
	\includegraphics[width=0.15\textwidth]{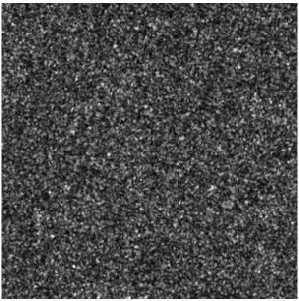}&
	\includegraphics[width=0.15\textwidth]{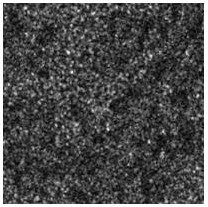}&
	\includegraphics[width=0.15\textwidth]{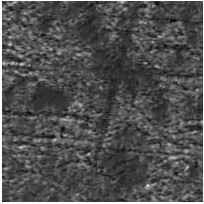}&
	\includegraphics[width=0.15\textwidth]{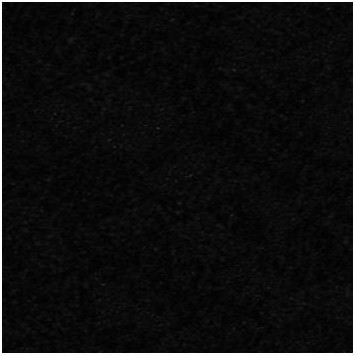}&
	\includegraphics[width=0.15\textwidth]{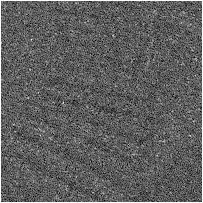}&
	\includegraphics[width=0.15\textwidth]{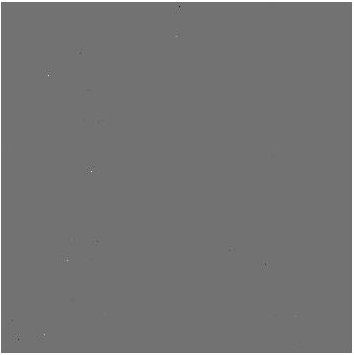}\\
		
\end{tabular}

\begin{tabular}{cccccc}
	\includegraphics[width=0.15\textwidth]{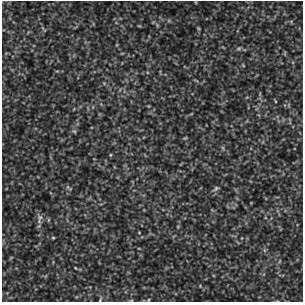}&
	\includegraphics[width=0.15\textwidth]{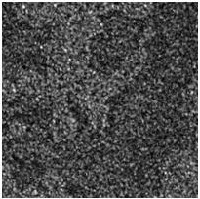}&
	\includegraphics[width=0.15\textwidth]{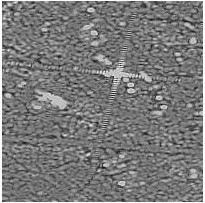}&
	\includegraphics[width=0.15\textwidth]{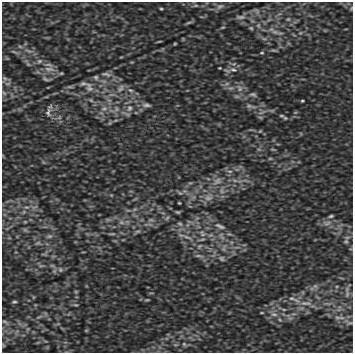}&
	\includegraphics[width=0.15\textwidth]{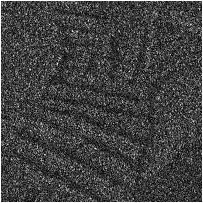}&
	\includegraphics[width=0.15\textwidth]{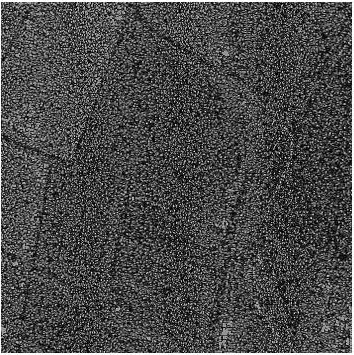}\\
		
\end{tabular}

	    \begin{tabular}{cccccc}
	    \includegraphics[width=0.15\textwidth]{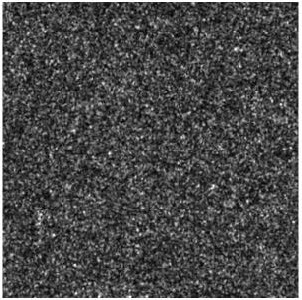}&
	    \includegraphics[width=0.15\textwidth]{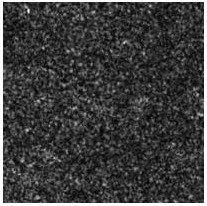}&
	    \includegraphics[width=0.15\textwidth]{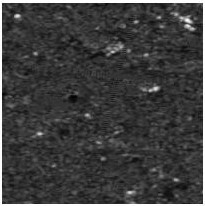}&
		\includegraphics[width=0.15\textwidth]{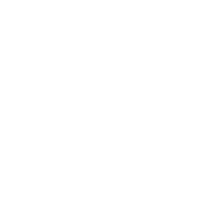}&
		\includegraphics[width=0.15\textwidth]{./Figures/NoData_speckle_red.png}&
		\includegraphics[width=0.15\textwidth]{./Figures/NoData_speckle_red.png}\\
	
	{(a)}&{(b)}& {(c)} &{(d)} &{(e)} &{(f)}
		
	\end{tabular}

	\caption{{Comparison of ratio images ($Y / \hat{X}$) for Improved Lee Sigma (top row), SAR-BM3D, SAR-CNN, deSpeckNet and reference ratio ($Y / {X}$ ) (bottom row)  (a) Indonesia, (b) Congo, (c) The Netherlands-Utrecht, (d) The Netherlands-Flevoland, (e) Japan, (f) Germany. {Since no clean image is available for (d)-(f), the last row is empty for those images.}} {To ease visual comparison, the ratio images were rescaled to the range [0.5 1.5].}   }\label{fig:Fig33}
\end{figure*}

\subsection{Tests on other sensors}
\noindent To further demonstrate the capability of deSpeckNet in generalizing to new sensors, we used the ALOS-2 PALSAR-2 image acquired in Japan.  Visually the performance of SAR-BM3D was better than the Indonesia and DRC images, due to sharper contrast and structure in the scene. However, there was a severe loss of resolution and texture in the image that resulted from over-filtering {and some spurious details were hallucinated by the filter as result of destruction of the texture in the image}. {The improved Lee sigma filter blurred all features achieving sub-optimal results and }SAR-CNN had a sub-optimal performance due to the difference in the noise distribution. In contrast, deSpeckNet was able to improve the SAR signal to noise ratio while preserving subtle features in the image (Figure~\ref{fig:Fig6}).  Same conclusions were reached when considering to the high resolution Iceye X2 image in Germany.  In this case also, deSpeckNet {performed better at} removing noise from homogeneous regions (Table~\ref{tbl:Tab4}) while preserving subtle features in the image (Figure~\ref{fig:Fig9}). \\

\subsection{Noise estimation}

\noindent One of the advantages of deSpeckNet is the estimation of the speckle noise distribution. This plays an important role in tuning the model to a different set of images.  This can be confirmed by investigating the probability density function of the estimated noise along with the parameters that define the noise probability density function. We do so by fitting a distribution (\ref{eq:2}) to the estimated noise image. As can be observed from (Figure~\ref{fig:Fig5}), the probability density function of the noise follows a Gamma distribution for all test areas except the Iceye Germany test area. Here, as opposed to the other test case we used a single look SAR image hence the speckle noise distribution was not a Gamma distribution as the other test cases but an exponential pdf and deSpeckNet was able to estimate the noise pdf accurately Figure~\ref{fig:Fig5}f. In addition, we also compared the reference image ratio ($Y/X$ with the ratio image ($Y/\hat{X}$) estimated by deSpeckNet {and the other methods (Figure~\ref{fig:Fig33}). {From Figure~\ref{fig:Fig33} we can clearly see that deSpeckNet inherently have a limitation in yielding uncorrelated speckle in the image, which is manifested by residual image structures in the ratio images. These artefacts in the ratio images are the result of using input images as a reference label in phase II to preserve features in the filtered output. To get a deeper insight to the performance of the methods} we used the mean of ratio (MoR) and variance of ratio (VoR) metric to compare the performance of the baseline methods.  In the Indonesia, DRC and Netherlands-Utrecht  images, deSpeckNet provides the closest estimate of MoR and VoR estimates to that of the reference ratio image dervived from the temporally averaged reference image $X$ (Table~\ref{tbl:Tab7}).} \\

\begin{table*}[tbh]
	\begin{center}
		\small
		\setlength\tabcolsep{12pt}
		\begin{tabular}{lccccc|c} \hline
	Test area & Metric & Lee Sigma & SAR-BM3D & SAR-CNN & deSpeckNet & Reference  \\ \hline
Indonesia & MoR &  {0.98} &  0.95 & 1.03 & {1.05}   & 1.05 \\
		  & VoR &  {0.13} &  0.03 & 0.15 & {0.16}  & 0.18 \\\hline
DRC       & MoR  & {0.98}&  0.95 & {1.01} & {0.93} &{1.02} \\ 
	      & VoR  & {0.15} &0.05 & 1.14 & {0.10} &{0.17} \\ \hline
NL-Utrecht & MoR  & {0.98}& {0.97} & 1.41 & {0.96} &  {0.95} \\ 
		   & VoR  & {0.04}& 0.01 & 0.07 & {0.02} & {0.04}  \\\hline
Japan      & MoR  & {0.99}& 0.89 & 0.93 & {0.96} &- \\  
           & VoR &  {0.23} &  0.16 & 0.07 & {0.23}  & -\\\hline
NL-Flevoland & MoR  & {0.97}& 0.96 & 0.91 & {0.83}  &- \\ 
            & VoR &  {0.11} &  0.02 & 6.37 & {0.18}  & -\\\hline
Germany    & Mor  & {0.98}& 0.78 & 1.11 & {1.04}&- \\ 
           & VoR &  {0.48} &  0.15 & 0.003 & {0.54}  & -\\\hline
			
		\end{tabular}\\
	\end{center}
	\caption{{Comparison of mean (MoR) and variance (VoR) for the ratio image ($Y / \hat{X}$) synthesized by the different methods and the reference ratio image ($Y / {X}$) that is synthesized from the temporally averaged reference image }}. \label{tbl:Tab7}
\end{table*}

\begin{figure*}
	\centering
	\begin{tabular}{ccccc}
		\includegraphics[width=0.2\textwidth]{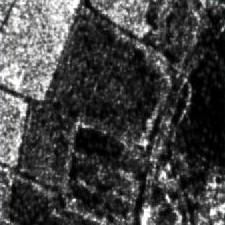}&
		\includegraphics[width=0.2\textwidth]{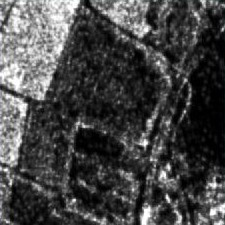}&
		\includegraphics[width=0.2\textwidth]{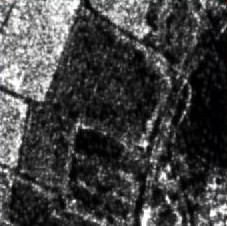}&
		\includegraphics[width=0.2\textwidth]{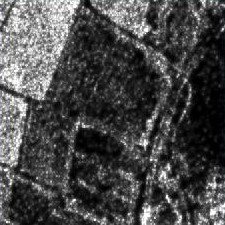}\\
		{(a)}&{(b)}& {(c)} &{(d)}
	\end{tabular}
	\caption{The effect of total variation loss in tuning a single look Iceye image. We show a $350 \times 350$ image patch for (a) Tuning with $\lambda = 10^{-3}, \mathrm{ENL} = 182.38$ (b) Tuning with $\lambda = 10^{-5}, \mathrm{ENL} = 165.7$ (c) Tuning with $\lambda = 10^{-6}, \mathrm{ENL} = 164.11$ (d) Tuning with $\lambda = 0, \mathrm{ENL} = 98.66$. Tuning is done in one epochs.}\label{fig:Fig10}
\end{figure*}

\subsection{Computational considerations}

\noindent The overall computational cost of deSpeckNet when training the initial model for 30 epochs was 31.8 hours. This was twice the computational burden of SAR-CNN, due to the duplicity of the CNN blocks in the Siamese architecture. In contrast, SAR-BM3D took 16 minutes to de-noise the input image in Indonesia. However, when tuning the model to new images the model was able to be tuned within 1 epoch for the DRC and Japan image and 2 epochs for the Netherlands image. This amounted to $7.7\times10^{-4}$ seconds per pixel to tune the model, as the computational burden depends on the dimension of the input image.  When testing the model on an image it had a computational burden of $2.69\times 10^{-5} $ seconds per pixel. All training was performed on GPU whereas, due to memory limitations, all testing was performed using CPU. \\

\begin{figure*}
	\centering
	\begin{tabular}{cccccc}
		\includegraphics[width=0.15\textwidth]{./Figures/Indo_despecknet_red.png}&
		\includegraphics[width=0.15\textwidth]{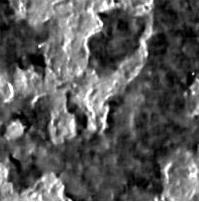}&
		\includegraphics[width=0.15\textwidth]{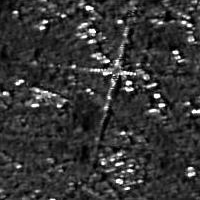}&
		\includegraphics[width=0.15\textwidth]{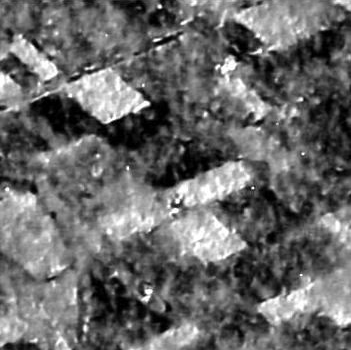}&
		\includegraphics[width=0.15\textwidth]{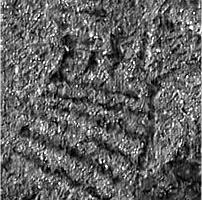}&
		\includegraphics[width=0.15\textwidth]{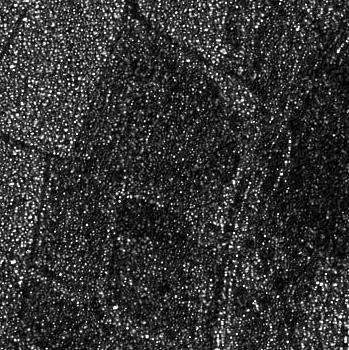}\\
		
	\end{tabular}
	
		\begin{tabular}{cccccc}
		\includegraphics[width=0.15\textwidth]{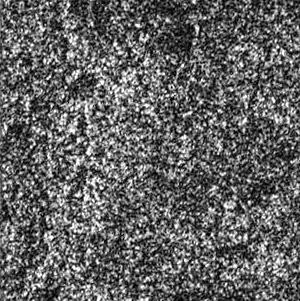}&
		\includegraphics[width=0.15\textwidth]{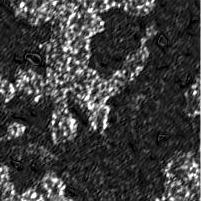}&
		\includegraphics[width=0.15\textwidth]{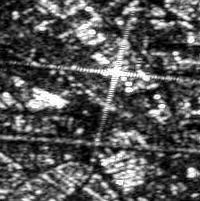}&
		\includegraphics[width=0.15\textwidth]{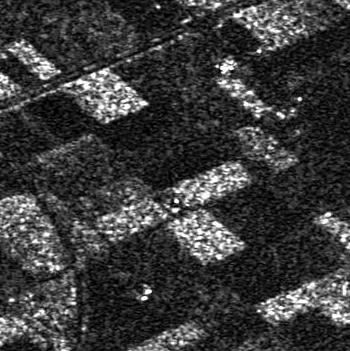}&
		\includegraphics[width=0.15\textwidth]{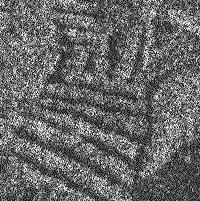}&
		\includegraphics[width=0.15\textwidth]{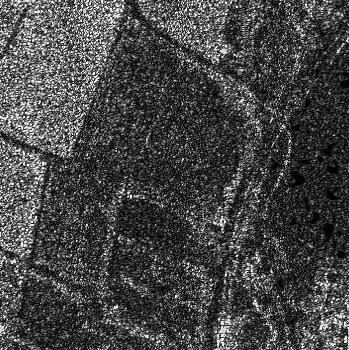}\\
	
	{(a)}&{(b)}& {(c)} &{(d)} &{(e)} &{(f)}
		
	\end{tabular}

	\caption{ {Comparison of processing output from phase one only (\textit{top}) and phase two only (\textit{bottom}). (a) Indonesia, (b) Congo, (c) The Netherlands-Utrecht, (d) The Netherlands-Flevoland, (e) Japan, (f) Germany.}  }\label{fig:Fig34}
\end{figure*}

\begin{table*}[tbh]
	\begin{center}
		\small
		\setlength\tabcolsep{8pt}
	{\begin{tabular}{llccccccc} \hline
			&Image & PSNR & SSIM & DG& EPI & ENL & Cx &$C_{NN}$\\ \hline
			\multirow{6}{*}{\rotatebox{90}{Phase 1 only}}&Indonesia & 45.76 &  0.97 & 11.51 &0.93 & 101.83 & 0.09 & -\\
			&DRC & 38.45 &  0.90 & 4.42& 0.65 & 26.03 & 0.19& -\\ 
			&NL-Utrecht & 18.80 & 0.63 & -8.39 & 0.76 & 62.90 & 0.12 & 2.50 \\
			&NL-Flevoland & - & -&  - & - & 14.98 &0.44 & -\\
			&Japan & - & -&   - & - & 10.71& 0.3&-\\ 
			&Germany & - & - & - & - & 3.92 &0.5 &- \\ \hline
			\multirow{6}{*}{\rotatebox{90}{Phase 2 only}}&Indonesia & 35.45 &  0.83 & 1.0 &  0.75 & 10.36 & 0.31& - \\
			&DRC & 35.32 &  0.84 & 1.27& 0.66 & 56.58 & 0.13 & -\\ 
			&NL-Utrecht & 24.39 & 0.69 & -3.26 & 0.52 & 1.78 & 0.74& 1.41  \\
			&NL-Flevoland & - &  - &  - & - & 12.83 & 0.27 & -\\
			&Japan & - & -&  - & - & 6.38 & & \\ 
			&Germany & - & -&- & - & 1.47 & & \\ \hline
			\multirow{6}{*}{\rotatebox{90}{Phase 1 and 2}}&Indonesia & 45.76 &  0.97 & 11.51 & 0.92 & 101.83 & 0.09 & - \\
			&DRC & 39.32 &  0.92 & 5.3 & 0.91 & 121.10 & 0.09 & - \\ 
			&NL-Utrecht & 27.60 & 0.83 & 0.94 & 0.84 & 242.61 & 0.062& 2.08 \\
			&NL-Flevoland & - &  - & - & - & 127.49 & 0.08 & -\\
			&Japan & - & -&  - & - & 158.21 & 0.09 & -\\ 
			&Germany & - & -& - & - & 65.22 & 0.12 & -\\ \hline
		\end{tabular}\\}
	\end{center}
	\caption{{Results of deSpeckNet, when using Phase 1 only (top block), Phase 2 only (middle block) and both Phase 1 and Phase 2 (proposed method, bottom block).}  }\label{tbl:Tab11} 
\end{table*}

\noindent Even though the computational cost of deSpeckNet was relatively heavy in the initial training phase, it was able to be tuned to new images with a relatively small amount of time. This makes deSpeckNet particularly desirable for operational applications that require fast response such as near-real time change monitoring applications. \\

\begin{table*}[tbh]
	\begin{center}
		\small
		\setlength\tabcolsep{20pt}
		\begin{tabular}{lcccccccc} \hline
			$\mu$ & $\lambda$ & $\xi$ & PSNR & SSIM & DG & {EPI}& ENL & Cx \\ \hline
			1 & 10$^{-2}$  & 10$^{-2}$ & 38.62 & 0.94& 4.59 & {\textbf{0.95}} & 62.65 & 0.12\\
			1 & 0  & 10$^{-2}$ & \textbf{45.76} & \textbf{0.97}& 11.51 & {0.93} & 101.83 & 0.09\\
			1 & 0  & 10$^{-1}$ & 45.66 & 0.97 & 11.48 & {0.91}& 111.61 & 0.09\\
			1 & 0  & 1 & 44.23 & 0.96 & 9.99 & {0.80} & \textbf{165.76} & \textbf{0.07} \\
			1 & 10$^{-3}$  & 10$^{-3}$ & 45.41 & 0.97 & 11.21 & {0.91} & 112.49 & 0.09\\
			1 & 0  & 10$^{-3}$ & 45.70 & 0.97 & 11.52 & {0.90} & 97.71 & 0.10\\
			1 & 10$^{-5}$  & 10$^{-3}$ & 45.64 & 0.97 & 11.45 & {0.94} & 112.83 & 0.09\\
			1 & 10$^{-7}$  & 10$^{-3}$ & 45.66 & 0.97 & 11.47 & {0.93} & 103.91 & 0.09\\ \hline
		\end{tabular}
	\end{center}
	\caption{PSNR, SSIM, {EPI} and ENL computed using different weights for the respective loss functions  for training the network using the image in Indonesia. Here $\mu$ is the weight of $L_{Clean}$, $\lambda$ is the weight of $L_{TV}$ and $\xi$ is the weight of $L_{Noisy}$. }\label{tbl:Tab12}
	
\end{table*}

\begin{table*}[tbh]
	\begin{center}
		\small
		\setlength\tabcolsep{20pt}
		\begin{tabular}{lccccccccc} \hline
			$\mu$ & $\lambda$ & $\xi$ & PSNR & SSIM & DG& {EPI} & ENL & Cx\\ \hline
			10$^{-2}$ & 10$^{-2}$  & 1 & 26.55 & 0.07 & -33.93 & {0.23} & 8.43 & -0.01\\
			10$^{-2}$ & 0  & 1 & \textbf{39.32} & \textbf{0.92} & \textbf{5.3} & {\textbf{0.92}}& 121.1 & \textbf{0.09} \\
			1.5 $\times$ 10$^{-2}$ & 10$^{-5}$  & 1 & 39.07 & 0.91 & 5.05 & {0.61} & 133.89 & 0.08\\
			1.5 $\times$ 10$^{-2}$ & 10$^{-4}$   & 1 & 38.65 & 0.91 & 4.63 & {0.63} & 70.85 & 0.11\\
			1.5 $\times$ 10$^{-2}$ & 10$^{-6}$  & 1 & 39.02 & 0.91 & 4.99& {0.61} & {147.66} & 0.08\\
			1.5 $\times$ 10$^{-3}$ & 10$^{-6}$  & 1 & 38.04 & 0.89 & 4.01 & {0.80} & \textbf{227.46} & 0.06 \\
			1.5 $\times$ 10$^{-1}$ & 10$^{-4}$  & 1 & 37.24 & 0.88 & 3.20& {0.89} & 33.92 & 0.17\\
			1.5  & 10$^{-3}$  & 1 & 36.90 & 0.87 & 2.86 & {0.78} & 42.84 &0.15 \\ \hline
		\end{tabular}
	\end{center}
	\caption{PSNR, SSIM, DG, {EPI}, ENL and Cx computed using different weights for the respective loss functions for tuning the network using the test image in the DRC. Here $\mu$ is the weight of $L_{Clean}$, $\lambda$ is the weight of $L_{TV}$ and $\xi$ is the weight of $L_{Noisy}$. 
	}\label{tbl:Tab13}
	
\end{table*}

\subsection{Ablation Study}
\noindent  {In this section, we} study the importance of the building blocks of the proposed architecture. {We first show the necessity of the two phases{, the training on the stack of multitemporal images (Phase 1) and the unsupervised fine tuning on the test image (phase 2), with respect to the full model using both (Table~\ref{tbl:Tab11} and Figure~\ref{fig:Fig34}).} From these results, its clear { that} the combination of the two phases is crucial to achieving higher performance in despeckling. Next, we performed a series of ablation studies focused on the loss functions.}  As shown in (Table~\ref{tbl:Tab12} and Table~\ref{tbl:Tab13}) the usage of a $L_{TV}$ in the initial training phase was not important to remove some artefacts and blurring effects found in the reconstructed image when the $L_{Clean}$ loss is applied (Table~\ref{tbl:Tab12}). However, when tuning the model {on noisier, single look images}, its presence was important to further smoothen noisy homogenous regions in the image to be tuned by forcing the $\mathrm{FCN_{noise}}$ part of the network in estimating the speckle component in the image. This is exemplified by the increase in the ENL when applying the total variation loss {on an Iceye single look} image (Figure~\ref{fig:Fig10}). To evaluate the necessity of applying a loss function in the $\mathrm{FCN_{clean}}$ side of the network, we removed both the $L_{Clean}$ and the $L_{TV}$ to train the network but it failed to achieve the output demonstrated when using, only  the $L_{Clean}$ loss. \\ 

\noindent In general, deSpeckNet achieved success in generalizing to different areas. It achieved higher success when applied to image in both rural and urban scene. This is attributed to the multiplicative model assumption enforced in deSpeckNet and the use of the input noisy image as a reference with a small weight.     \\

\section{Conclusion}

\noindent {We have presented a method, deSpeckNet, that is able to learn a speckle noise model suitable for effective despeckling without the need of any reference clean image \diego{at test time}{} nor any assumptions on the noise distribution {other than the multiplicative noise model}. Our experiments on a wide variety of SAR images, obtained with different sensors and over different regions, confirm the robustness of deSpeckNet.}\\

\noindent The proposed deSpeckNet proved to be effective in reducing speckle noise while preserving the image quality with minimal unsupervised fine tuning. It was also able to adapt to {all the tested} SAR images regardless of resolution, acquisition parameters or geographic region, providing better despeckling results than state-of-the-art methods {and equalling performance obtained by CNN models optimized with temporally averaged images, generally unavailable, at test time}.  For future work, we plan on improving the loss functions {that encode the assumption on the clean image} to improve the performance of deSpeckNet in mixed urban and rural scenes where both strong deterministic and distributed targets exist.\\


\section*{Acknowledgement}

\noindent This work was partly funded through the Global Forest Watch-World Resources Institute (GFW-WRI) Radar for Detecting Deforestation (RADD) project and the U.S. government SilvaCarbon Program.

\bibliographystyle{unsrt}
\bibliography{References.bib}

\begin{thebibliography}{10}

\bibitem{porcello1976speckle}
Leonard~J Porcello, Norman~G Massey, Richard~B Innes, and James~M Marks.
\newblock Speckle reduction in synthetic-aperture radars.
\newblock {\em JOSA}, 66(11):1305--1311, 1976.

\bibitem{lee1980digital}
Jong-Sen Lee.
\newblock Digital image enhancement and noise filtering by use of local
  statistics.
\newblock {\em IEEE Transactions on Pattern Analysis \& Machine Intelligence},
  (2):165--168, 1980.

\bibitem{lee1994intensity}
Jong-Sen Lee, Karl~W Hoppel, Stephen~A Mango, and Allen~R Miller.
\newblock Intensity and phase statistics of multilook polarimetric and
  interferometric {SAR} imagery.
\newblock {\em IEEE Transactions on Geoscience and Remote Sensing},
  32(5):1017--1028, 1994.

\bibitem{frost1982model}
Victor~S Frost, Josephine~Abbott Stiles, K~Sam Shanmugan, and Julian~C
  Holtzman.
\newblock A model for radar images and its application to adaptive digital
  filtering of multiplicative noise.
\newblock {\em IEEE Transactions on Pattern Analysis and Machine Intelligence},
  (2):157--166, 1982.

\bibitem{lee1999polarimetric}
Jong-Sen Lee, Mitchell~R Grunes, and Gianfranco De~Grandi.
\newblock Polarimetric {SAR} speckle filtering and its implication for
  classification.
\newblock {\em IEEE Transactions on Geoscience and Remote Sensing},
  37(5):2363--2373, 1999.

\bibitem{vasile2006intensity}
Gabriel Vasile, Emmanuel Trouv{\'e}, Jong-Sen Lee, and Vasile Buzuloiu.
\newblock Intensity-driven adaptive-neighborhood technique for polarimetric and
  interferometric {SAR} parameters estimation.
\newblock {\em IEEE Transactions on Geoscience and Remote Sensing},
  44(6):1609--1621, 2006.

\bibitem{lee2005scattering}
Jong-Sen Lee, Mitchell~R Grunes, Dale~L Schuler, Eric Pottier, and Laurent
  Ferro-Famil.
\newblock Scattering-model-based speckle filtering of polarimetric {SAR} data.
\newblock {\em IEEE Transactions on Geoscience and Remote Sensing},
  44(1):176--187, 2005.

\bibitem{deledalle2014nl}
Charles-Alban Deledalle, Lo{\"\i}c Denis, Florence Tupin, Andreas Reigber, and
  Marc J{\"a}ger.
\newblock Nl-sar: A unified nonlocal framework for resolution-preserving
  (pol)(in) {SAR} denoising.
\newblock {\em IEEE Transactions on Geoscience and Remote Sensing},
  53(4):2021--2038, 2014.

\bibitem{dabov2007image}
Kostadin Dabov, Alessandro Foi, Vladimir Katkovnik, and Karen Egiazarian.
\newblock Image denoising by sparse 3-d transform-domain collaborative
  filtering.
\newblock {\em IEEE Transactions on Image Processing}, 16(8):2080--2095, 2007.

\bibitem{molina2011evaluation}
Daniela~Espinoza Molina, Du{\v{s}}an Gleich, and Mihai Datcu.
\newblock Evaluation of bayesian despeckling and texture extraction methods
  based on gauss--markov and auto-binomial gibbs random fields: application to
  terra{SAR}-x data.
\newblock {\em IEEE Transactions on Geoscience and Remote Sensing},
  50(5):2001--2025, 2011.

\bibitem{mahdianpari2017effect}
Masoud Mahdianpari, Bahram Salehi, and Fariba Mohammadimanesh.
\newblock The effect of {PolSAR} image de-speckling on wetland classification:
  introducing a new adaptive method.
\newblock {\em Canadian Journal of Remote Sensing}, 43(5):485--503, 2017.

\bibitem{lopes1990maximum}
Armand Lopes, E~Nezry, R~Touzi, and H~Laur.
\newblock Maximum a posteriori speckle filtering and first order texture models
  in {SAR} images.
\newblock In {\em 10th annual International Symposium on Geoscience and Remote
  Sensing}, pages 2409--2412. IEEE, 1990.

\bibitem{chierchia2017sar}
Giovanni Chierchia, Davide Cozzolino, Giovanni Poggi, and Luisa Verdoliva.
\newblock {SAR} image despeckling through convolutional neural networks.
\newblock In {\em 2017 IEEE International Geoscience and Remote Sensing
  Symposium (IGARSS)}, pages 5438--5441. IEEE, 2017.

\bibitem{zhang2017beyond}
Kai Zhang, Wangmeng Zuo, Yunjin Chen, Deyu Meng, and Lei Zhang.
\newblock Beyond a gaussian denoiser: Residual learning of deep {CNN} for image
  denoising.
\newblock {\em IEEE Transactions on Image Processing}, 26(7):3142--3155, 2017.

\bibitem{zhang2018learning}
Qiang Zhang, Qiangqiang Yuan, Jie Li, Zhen Yang, and Xiaoshuang Ma.
\newblock Learning a dilated residual network for {SAR} image despeckling.
\newblock {\em Remote Sensing}, 10(2):196, 2018.

\bibitem{vitale2019new}
Sergio Vitale, Giampaolo Ferraioli, and Vito Pascazio.
\newblock A new ratio image based {CNN} algorithm for {SAR} despeckling.
\newblock {\em arXiv preprint arXiv:1906.04111}, 2019.

\bibitem{pan2019filter}
Ting Pan, Dong Peng, Wen Yang, and Heng-Chao Li.
\newblock A filter for {SAR} image despeckling using pre-trained convolutional
  neural network model.
\newblock {\em Remote Sensing}, 11(20):2379, 2019.

\bibitem{deledalle2017mulog}
Charles-Alban Deledalle, Lo{\"\i}c Denis, Sonia Tabti, and Florence Tupin.
\newblock Mu{L}o{G}, or how to apply gaussian denoisers to multi-channel {SAR}
  speckle reduction?
\newblock {\em IEEE Transactions on Image Processing}, 26(9):4389--4403, 2017.

\bibitem{lattari2019deep}
Francesco Lattari, Borja Gonzalez~Leon, Francesco Asaro, Alessio Rucci, Claudio
  Prati, and Matteo Matteucci.
\newblock Deep learning for {SAR} image despeckling.
\newblock {\em Remote Sensing}, 11(13):1532, 2019.

\bibitem{ulaby2014microwave}
Fawwaz~Tayssir Ulaby, David~G Long, William~J Blackwell, Charles Elachi,
  Adrian~K Fung, Chris Ruf, Kamal Sarabandi, Howard~A Zebker, and Jakob
  Van~Zyl.
\newblock {\em Microwave Radar and Radiometric Remote Sensing}, volume~4.

\bibitem{mullissa2019polsarnet}
Adugna~G Mullissa, Claudio Persello, and Alfred Stein.
\newblock {P}ol{SARN}et: A deep fully convolutional network for polarimetric
  {SAR} image classification.
\newblock {\em IEEE Journal of Selected Topics in Applied Earth Observations
  and Remote Sensing}, 2019.

\bibitem{ioffe2015batch}
Sergey Ioffe and Christian Szegedy.
\newblock Batch normalization: Accelerating deep network training by reducing
  internal covariate shift.
\newblock In {\em International Conference on Machine Learning}, pages
  448--456, 2015.

\bibitem{dahl2013improving}
George~E Dahl, Tara~N Sainath, and Geoffrey~E Hinton.
\newblock Improving deep neural networks for {LVCSR} using rectified linear
  units and dropout.
\newblock In {\em 2013 IEEE international Conference on Acoustics, Speech and
  Signal Processing}, pages 8609--8613. IEEE, 2013.

\bibitem{aubert2008variational}
Gilles Aubert and Jean-Francois Aujol.
\newblock A variational approach to removing multiplicative noise.
\newblock {\em SIAM Journal on Applied Mathematics}, 68(4):925--946, 2008.

\bibitem{zhao2019ratio}
Weiying Zhao, Charles-Alban Deledalle, Lo{\"\i}c Denis, Henri Ma{\^\i}tre,
  Jean-Marie Nicolas, and Florence Tupin.
\newblock Ratio-based multitemporal sar images denoising: Rabasar.
\newblock {\em IEEE Transactions on Geoscience and Remote Sensing},
  57(6):3552--3565, 2019.

\bibitem{kingma2014adam}
Diederik~P Kingma and Jimmy Ba.
\newblock Adam: A method for stochastic optimization.
\newblock {\em arXiv preprint arXiv:1412.6980}, 2014.

\bibitem{glorot2010understanding}
Xavier Glorot and Yoshua Bengio.
\newblock Understanding the difficulty of training deep feedforward neural
  networks.
\newblock In {\em Proceedings of the {T}hirteenth {I}nternational {C}onference
  on {A}rtificial {I}ntelligence and {S}tatistics}, pages 249--256, 2010.

\bibitem{vedaldi2015matconvnet}
Andrea Vedaldi and Karel Lenc.
\newblock Matconvnet: Convolutional neural networks for {M}atlab.
\newblock In {\em Proceedings of the 23rd ACM International Conference on
  Multimedia}, pages 689--692. ACM, 2015.

\bibitem{di2013benchmarking}
Gerardo Di~Martino, Mariana Poderico, Giovanni Poggi, Daniele Riccio, and Luisa
  Verdoliva.
\newblock Benchmarking framework for {SAR} despeckling.
\newblock {\em IEEE Transactions on Geoscience and Remote Sensing},
  52(3):1596--1615, 2013.

\bibitem{mullissa2017scattering}
Adugna~G. Mullissa, Valentyn Tolpekin, and Alfred Stein.
\newblock Scattering property based contextual {P}ol{SAR} speckle filter.
\newblock {\em International Journal of Applied Earth Observation and
  Geoinformation}, 63:78--89, 2017.

\bibitem{foucher2014analysis}
Samuel Foucher and Carlos L{\'o}pez-Mart{\'\i}nez.
\newblock Analysis, evaluation, and comparison of polarimetric {SAR} speckle
  filtering techniques.
\newblock {\em IEEE Transactions on Image Processing}, 23(4):1751--1764, 2014.

\bibitem{lee2008improved}
Jong-Sen Lee, Jen-Hung Wen, Thomas~L Ainsworth, Kun-Shan Chen, and Abel~J Chen.
\newblock Improved sigma filter for speckle filtering of {SAR} imagery.
\newblock {\em IEEE Transactions on Geoscience and Remote Sensing},
  47(1):202--213, 2008.

\end{thebibliography}

\end{document}